\documentclass[aps,prc,twocolumn,groupedaddress]{revtex4-1}
\usepackage{graphicx}
\usepackage{subfigure}
\usepackage{hyperref}
\usepackage[nameinlink]{cleveref}
\usepackage{array}
\usepackage{multirow}
\usepackage{soul}

\begin{document}

\title{Color reconnection as a possible mechanism of intermittency in the emission spectra of charged particles in PYTHIA-generated high-multiplicity $\textit{pp}$ collisions at energies available at the CERN Large Hadron Collider}

\author{Pranjal Sarma}
\email[]{pranjal.sarma@cern.ch}
\author{Buddhadeb Bhattacharjee}
\email[Corresponding author: ]{buddhadeb.bhattacharjee@cern.ch}
\affiliation{Nuclear and Radiation Physics Research Laboratory, Department of Physics, Gauhati University, Guwahati, Assam - 781014, India}

\date{\today}

\begin{abstract}

Nonstatistical fluctuation in pseudorapidity ($\eta$), azimuthal ($\phi$), and pseudorapidity-azimuthal ($\eta-\phi$) distribution spectra of primary particles of PYTHIA Monash (default) generated $pp$ events at $\sqrt{s}=$ 2.76, 7, and 13 TeV have been studied using the scaled factorial moments technique. A weak intermittent type of emission could be realized for minimum-bias (MB) $pp$ events in $\chi(\eta-\phi)$ space and a much stronger intermittency could be observed in high-multiplicity (HM) $pp$ events in all $\chi(\eta)$, $\chi(\phi)$, and $\chi(\eta-\phi)$ spaces at all the studied energies. For HM $pp$ events, at a particular energy, the intermittency index $\alpha_{q}$ is found to be largest in two-dimensional $\chi(\eta-\phi)$ space and least in $\chi(\eta)$ space, and no center of mass energy dependence of $\alpha_{q}$ could be observed. The anomalous dimensions $d_{q}$ are observed to be increased with the order of the moment $q$, suggesting a multifractal nature of the emission spectra of various studied events. While, the coefficient $\lambda_q$ is found to decrease monotonically with the order of the moment $q$ for two-dimensional analysis of MB $pp$ events as well as for one-dimensional analysis of HM $pp$ events, a clear minimum in $\lambda_q$ values could be observed from the two-dimensional HM $pp$ data analysis. For PYTHIA Monash generated sets of data, the strength of the intermittency is found to vary significantly with the variation of the strength of the color reconnection (CR) parameter, i.e., reconnection range RR, for RR = 0.0, 1.8 and 3.0, thereby, establishing a strong connection between the CR mechanism and the observed intermittent type of emission of primary charged particles of the studied high-multiplicity $pp$ events.
\end{abstract}



\maketitle

\section{Introduction} 

Understanding the particle production mechanism is one of the primary goals of high-energy $A+A$ and $h+h$ collisions. A characteristic feature of primary charged particles produced in any such collision is that they exhibit fluctuation in particle number densities over the pseudorapidity space. Such fluctuation is much larger than the statistical fluctuations arising due to the finiteness of the yield of particles produced in a collision. In the pseudorapidity distribution spectra, these fluctuations manifest themselves as peaks and valleys in narrow domains of pseudorapidity space. Such anomalous fluctuation resulting in a "spike" like structure in single particle density distribution spectrum is often used to examine if the nuclear matter has undergone a phase transition during the evolution of a collision \cite{vhove, gyulassy, hwa}. Further, a study on such a fluctuation is also important from the point of view that such anomalous spatial fluctuations may arise due to many mini-jets that might have been formed as a result of semi-hard parton-parton interactions or gluon bremsstrahlung \cite{bialas2, ochs1, bialas1}.

To gather any meaningful information about the particle production mechanism, it is therefore important to disentangle and analyze these anomalous fluctuations, arising out of some dynamical processes, from that of noise arising due to the finite number of available particles in the final state. The scaled factorial moment (SFM) technique, as proposed by Bialas and Peschanski \cite{bialas2}, is found to be a useful mathematical tool that separates the dynamical fluctuation from the mixture of the two. According to this prescription, a power law growth of the averaged scaled factorial moments ($\langle F_{q} \rangle$) with the decrease of the phase space bin width ($\delta w$), or otherwise, the number of bins $M$ into which the entire phase space is divided, that is, $\langle F_{q} \rangle \propto (M)^{\alpha_{q}}$ is referred to as intermittency and thus indicates the presence of the contribution of the dynamical fluctuation in the data sample. Intermittency, in turn, is found to be related to self-similarity and fractality of emission spectra as well as the particle emitting source \cite{sarcevic}. The exponent $\alpha_{q}$ of the power law, called the intermittency index, is connected with the anomalous dimension $d_{q}$ ($= D-D_{q}$) through the relation $d_{q}=\alpha_{q}/(q-1)$, where $D$ is the ordinary topological dimension of the space into which the fractal objects are embedded and $D_{q}$ is the generalized $q^{th}$ order Renyi dimension \cite{lipa, bhattacharjee1}. Knowledge of order dependence of $d_{q}$ is helpful to make comments on the fractal nature of emission spectra and in turn on particle production mechanism and associated phase transition, if any.

Indications of the existence of a new state of deconfined matter, called quark-gluon plasma (QGP) have been provided for $A+A$ collisions by previous studies at the CERN Super Proton Synchrotron (SPS) \cite{heinz} and at the Relativistic Heavy-Ion Collider (RHIC) \cite{arsene, adcox, bback, jadams_star, shuryak, huovinen, muller} and then at the CERN Large Hadron Collider (LHC) \cite{abelev1, jadam_alice1, jadam_alice2, abelev2}. On the other hand, in small systems like $pp$ collisions, with a few available partons for the collisions, no such phase transition is expected and is traditionally considered as the reference system for the heavy-ion collision studies. Due to early thermalization, it also remains doubtful if the matter formed in such collisions exhibits collective-like behavior as observed in heavy-ion collisions \cite{bzdak}. However, with the increased energy for the proton-proton colliding system at the LHC, the high-multiplicity $pp$ events reach multiplicity comparable to proton-nucleus and nucleus-nucleus collisions \cite{jadam_alice3}. Recent experimental measurements in $pp$ collisions show that the flow-like effects do exist in high-multiplicity $pp$ events as well giving an indication of collective behavior in such a small system \cite{jadam_alice3}. Thus, systematic studies on various observable of high-multiplicity $pp$ events have become essential for a better understanding of the dynamics of such collisions.

The PYTHIA Monte Carlo event generator is a general purpose perturbative-QCD based event generator. It uses a factorized perturbative expansion for the hard parton-parton interaction, combined with parton showers, and details models for hadronization and multiple parton interactions. It has been extensively used to describe $pp$ collisions data and is found to be quite successful in describing the various experimental results of $pp$ collisions at the LHC energies \cite{acharya_alice1, jadam_alice4, abelev4}. Different parameters of the existing PYTHIA model have been improved or tuned from time to time to describe the data well. In the PYTHIA Monash version, the parameters are tuned in such a way that a better description of the experimental data at the LHC energies $\sqrt{s}=$ 7 and 13 TeV could be achieved \cite{skands}. Further details of PYTHIA event generator are available in Refs. \cite{skands, pythia_6}. In this work, an attempt has been made with PYTHIA Monash generated data to study, within ALICE acceptance, the nonstatistical fluctuation of single-particle density distribution spectrum in the light of scaled factorial moments (SFMs) for $pp$ collisions at the LHC energies $\sqrt{s}=$ 2.76, 7, and 13 TeV.

Further, in PYTHIA, color reconnection (CR), a string fragmentation model, has been implemented where the final partons are considered to be color connected in such a way that the total string length becomes as short as possible \citep{gustafson}. It has been reported recently that the CR mechanism in PYTHIA can mimic the effect of collective-like behavior, such as a mass-dependent rise in $\langle p_{T} \rangle$ with multiplicity, a bump in baryon to meson ratio at intermediate $p_{T}$, etc. as observed in heavy-ion collisions \citep{antonio, abelev1}. Such behavior is attributed to the fact that due to CR, partons from two independent interactions select a preferred pseudorapidity ($\eta$) and azimuthal ($\phi$) angle of emissions giving rise to a boost in $p_{T}$ and higher-order flow.  Since, an increase in scaled factorial moment with the decrease of phase-space bin width is considered to be an indication of the presence of large fluctuations in the data sample, which in turn is related to the phase transition and particle production mechanism, in this work, an attempt has also been made to find if color reconnection has any significant role to play on intermittency and other related observables of nuclear collisions.

\section{The Scaled Factorial Moments Technique}	

Even though the technique of estimation of the scaled factorial moment is a well-established mathematical tool of high-energy nuclear collision studies and is described in details in a number of works \cite{bialas2, ochs1, bialas1, bhattacharjee1, zhang, mali, xie, swarna}, for completeness of the article only the relevant steps will be described hereunder.

Let us consider the distribution of the charged particles in the pseudorapidity [$\eta = -$ln $tan(\theta/2)$] space. Let $\Delta \eta$ be the overall interval of the space, and $n$ is the total number of particles in an event within $\Delta \eta$. If the overall interval is divided into $M$ equal parts, we get bins of smaller width $\delta \eta=\Delta \eta/M$, and if the number of particles falling within the $m$th such bin is $n_{m}$, then $n=\sum{n_{m}}$, where the summation runs over m from $m = 1$ to $M$. The factorial moment ($f_{q}$) of order $q$ can now be calculated from the $m$th bin as \cite{bhattacharjee2}:

\begin{equation}
f_{q}=n_{m}(n_{m}-1)\cdots (n_{m}-q+1)
\label{f_q_moment}
\end{equation}

Such moments for a particular bin are first calculated for all events and then averaged over all events. This average moment is then calculated for all bins and again averaged over all bins. This averaging procedure is called vertical averaging \cite{bialas2, xie, bhattacharjee2}.

On the other hand, if the factorial moments is first calculated for a bin and then averaged over all bins and this averaged moments is then calculated for all events and again averaged over all events, then  this method of estimation of scaled factorial moments is called horizontal averaging \cite{bialas1, bialas2, bhattacharjee1, xie}. While the method of vertical averaging takes accounts of the dynamical fluctuation in event space, the horizontal averaging takes account of the non-statistical fluctuation in phase space of an event.

The expression for the vertically averaged scaled factorial moment is given by:

\begin{equation}
\langle F_{q} \rangle = \frac{1}{M} \sum_{m=1}^{M}\frac{1}{N} \sum_{i=1}^{N} \frac{n_{m}(n_{m}-1) \cdots  (n_{m}-q+1)}{\langle n_{m}\rangle^{q}}
\label{VHSM_equation}
\end{equation}

and the horizontally averaged scaled factorial moments is expressed as:

\begin{equation}
\langle F_{q}\rangle=\frac{1}{N} \sum_{i=1}^{N}M^{q-1} \sum_{m=1}^{M} \frac{n_{m}(n_{m}-1) \cdots (n_{m}-q+1)}{\langle n\rangle^{q}}
\label{HSFM_equation}
\end{equation}

where, $N$ is the total number of events in the data sample.

\begin{figure*}[htp]
\subfigure[Minimum bias pp collisions]{\includegraphics[width=70mm, height=52mm]{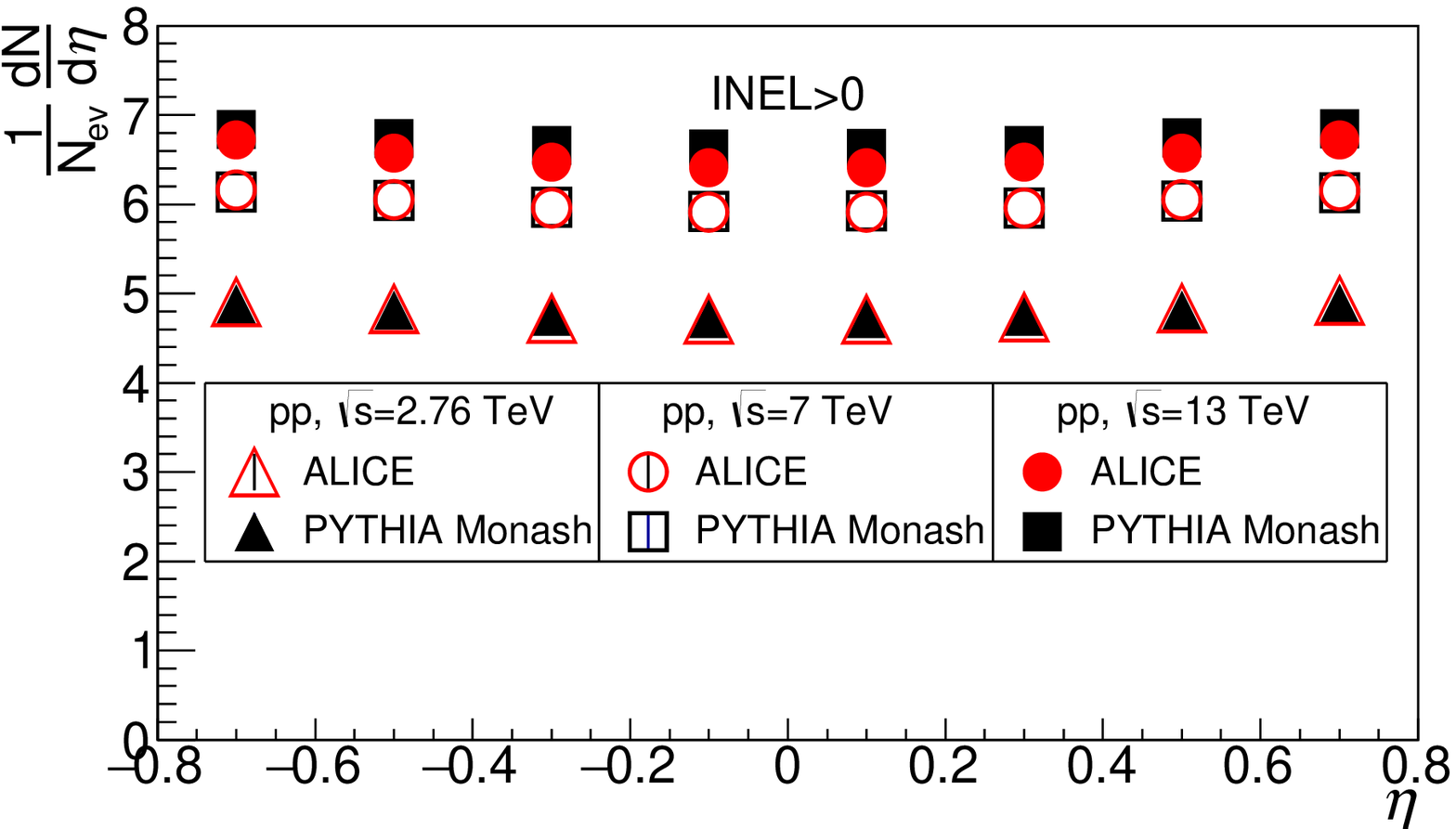}\label{eta_dist}}
\subfigure[High multiplicity pp collisions]{\includegraphics[width=70mm, height=52mm]{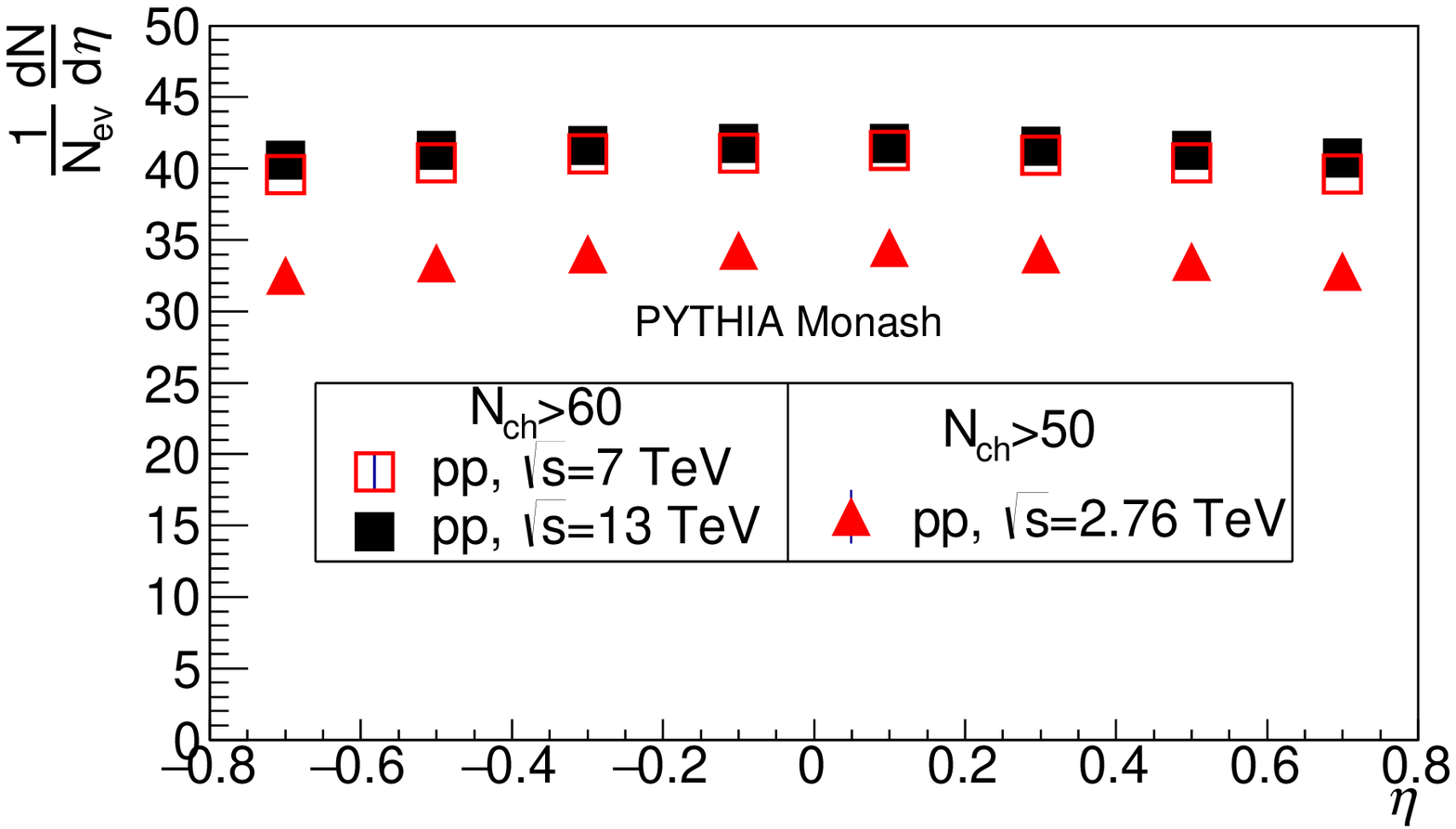}\label{eta_hm}}
\caption{(Color online) Pseudorapidity distribution of the primary charged particles for PYTHIA Monash (default) generated data (a) compared with experimental data of ALICE in minimum-bias $pp$ collisions \cite{jadam_alice4, jadam_alice5} and (b) in high-multiplicity $pp$ events at $\sqrt{s}$ = 2.76, 7, and  13 TeV.}
\end{figure*}

Here, it is worth mentioning that the horizontal scaled factorial moment technique has the limitation of its dependence on the shape of the single particle density distribution spectra. However, the shape dependence of the horizontally averaged SFM can be eliminated by converting the distribution of the particles in pseudorapidity space to a distribution of a new cumulative variable $\chi(\eta)$, defined as \cite{bialas3, ghosh, adamovich, wang}:

\begin{equation}
\chi(\eta)=\frac{\int_{\eta_{min}}^{\eta} \rho(\eta)d\eta}{\int_{\eta_{max}}^{\eta_{min}} \rho(\eta)d\eta}
\label{chi_equation}
\end{equation}

In $\chi(\eta)$ space, the density distribution spectrum would be perfectly flat.

For power law type dependence of $\langle F_{q} \rangle$ on $M$ i.e. if $\langle F_{q} \rangle \propto (M)^{\alpha_{q}}$, a plot of ln$\langle F_{q} \rangle$ vs. ln$M$ should be a straight line with the slope as the exponent of the power law.

\begin{equation}
\alpha_{q}=\frac{\Delta ln \langle F_{q} \rangle}{\Delta lnM}
\end{equation}

$\alpha_{q}$, called the intermittency index, is related to the anomalous fractal dimension $d_{q}$ ($=D-D_{q}$) through the relation:

\begin{equation}
d_{q}=\frac{\alpha_{q}}{q-1}
\label{dq_equation}
\end{equation}

An order invariance of $d_{q}$ refers to monofractality whereas an increase of $d_{q}$ with $q$ refers to the multifractal nature of the emission spectra \cite{lipa, bhattacharjee1}.

\section{Results and discussions}

A total of approximately 189 $\times$ $10^6$ million events were generated for $pp$ collisions at $\sqrt{s}=$ 2.76, 7, and 13 TeV using PYTHIA Monash (default) Monte Carlo (MC) event generator.

The pseudorapidity distributions of the primary charged particles within the acceptance of ALICE detector, for both minimum-bias (MB) and high-multiplicity ($N_{ch}>50$ for 2.76 TeV and $N_{ch}>60$ for 7 and 13 TeV) (HM) $pp$ collisions at $\sqrt{s}=$ 2.76, 7, and 13 TeV with the present set of generated data are shown in Figs.~\ref{eta_dist} and ~\ref{eta_hm} respectively and compared with the existing experimental results of ALICE Collaboration for inelastic (INEL) $>$0 \cite{jadam_alice4, jadam_alice5}. From these figures, it is found that the pseudorapidity distribution of MC events agrees well with the experimental data for the studied region of $|\eta| < 0.8$. It is therefore believed that further analysis of our generated data using the scaled factorial moment technique might be of some significance.

\begin{figure}[h]
\includegraphics[width=68mm, height=51mm]{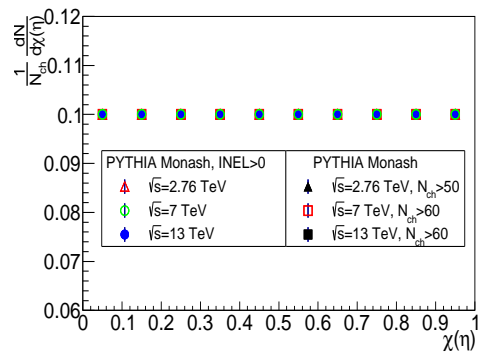}
\caption{(Color online) $\chi(\eta)$ distribution of the primary charged particles produced in MB and HM $pp$ events at $\sqrt{s}$ = 2.76, 7, and 13 TeV with PYTHIA Monash (default) generated data.}
\label{chi_dist}
\end{figure}

Figure ~\ref{chi_dist} represents the same distributions of Figs.~\ref{eta_dist} and ~\ref{eta_hm}, but in $\chi(\eta)$ space for the generated sets of data only for MB and HM $pp$ events of $\sqrt{s}=$ 2.76, 7, and 13 TeV. It could be readily seen from Fig.~\ref{chi_dist} that, as expected, the various distributions of Figs.~\ref{eta_dist} and ~\ref{eta_hm} become perfectly flat in $\chi(\eta)$ space.

\begin{figure*}[ht]
\subfigure[]{\includegraphics[width=60mm, height=50mm]{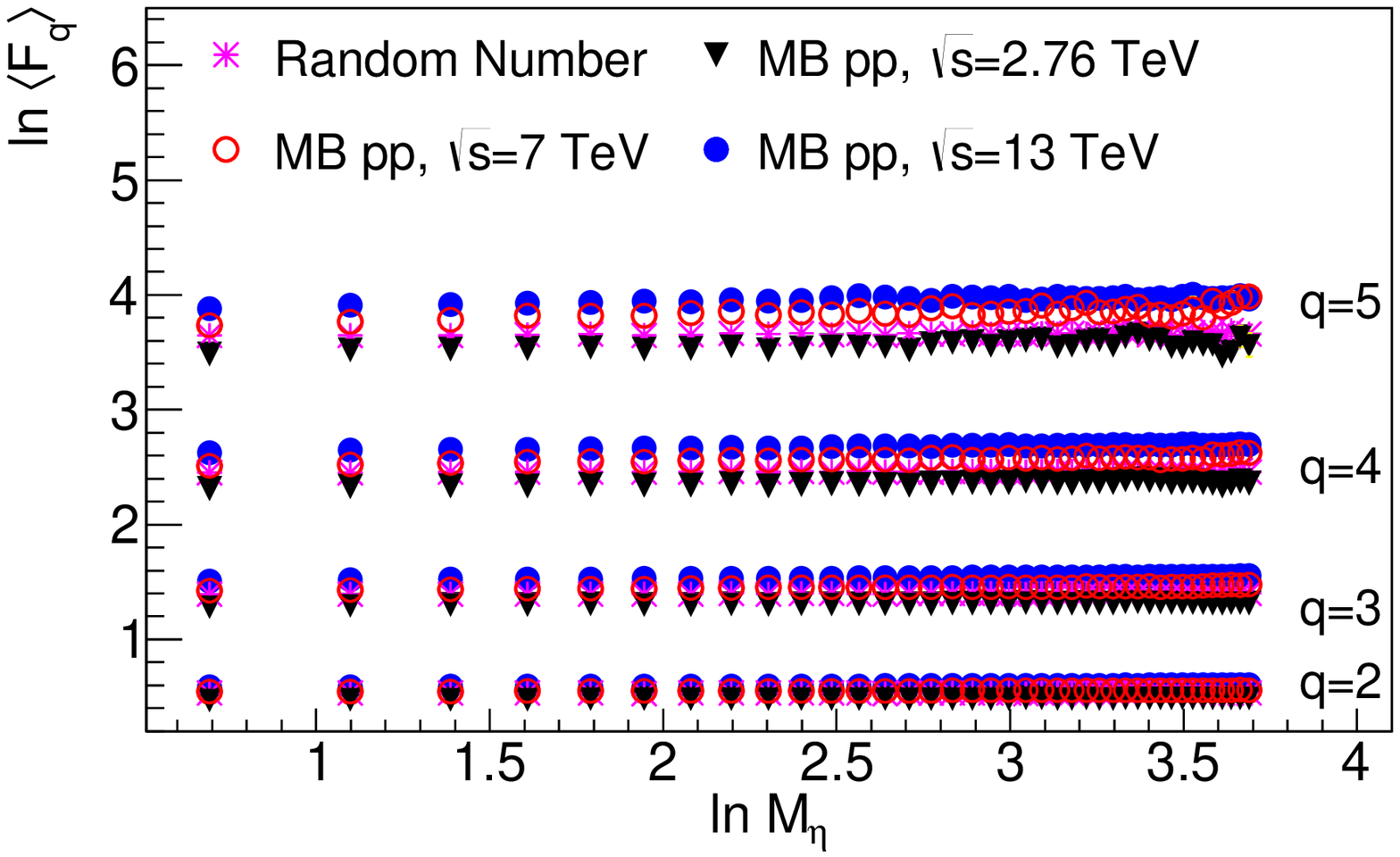}\label{sfm_mb190}}
\subfigure[]{\includegraphics[width=118mm, height=50mm]{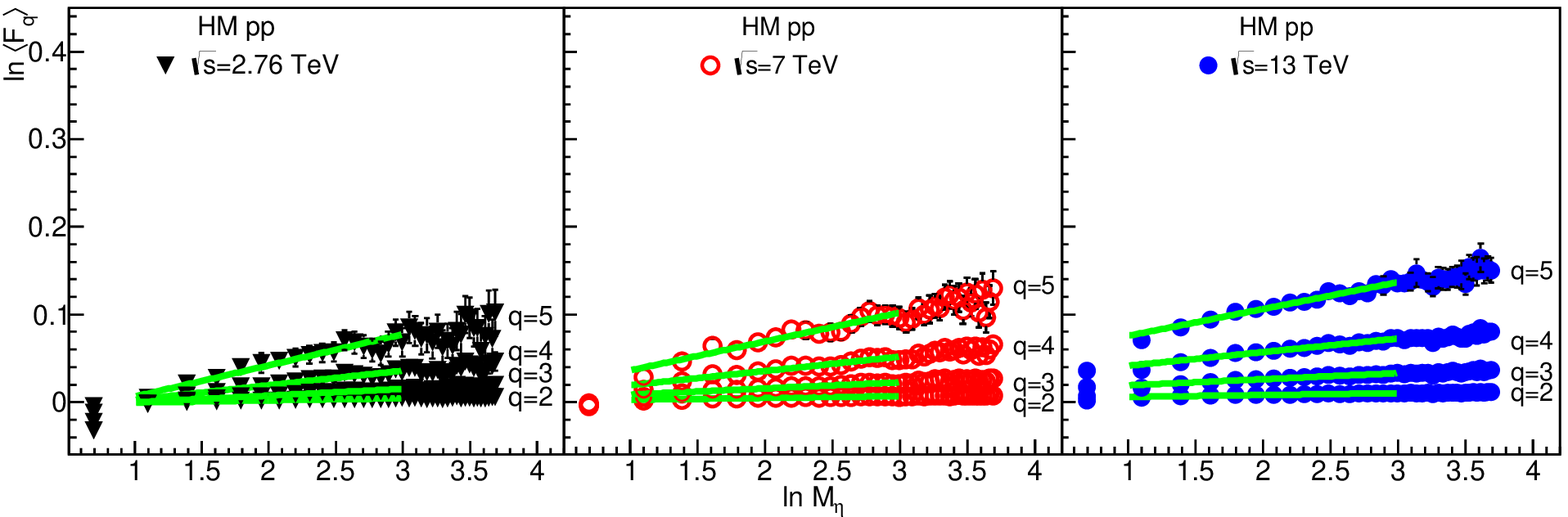}\label{sfm_hm190}}
\caption{(Color online) ln$\langle F_{q} \rangle$ vs. ln$M$ for moments $q=2-5$ for (a) RAN, MB $pp$ collisions and (b) HM $pp$ collisions at $\sqrt{s}=$ 2.76, 7, and 13 TeV in one-dimensional $\chi(\eta)$ space with PYTHIA Monash (default) generated data.}
\end{figure*}

\begin{figure*}[ht]
\subfigure[]{\includegraphics[width=60mm, height=50mm]{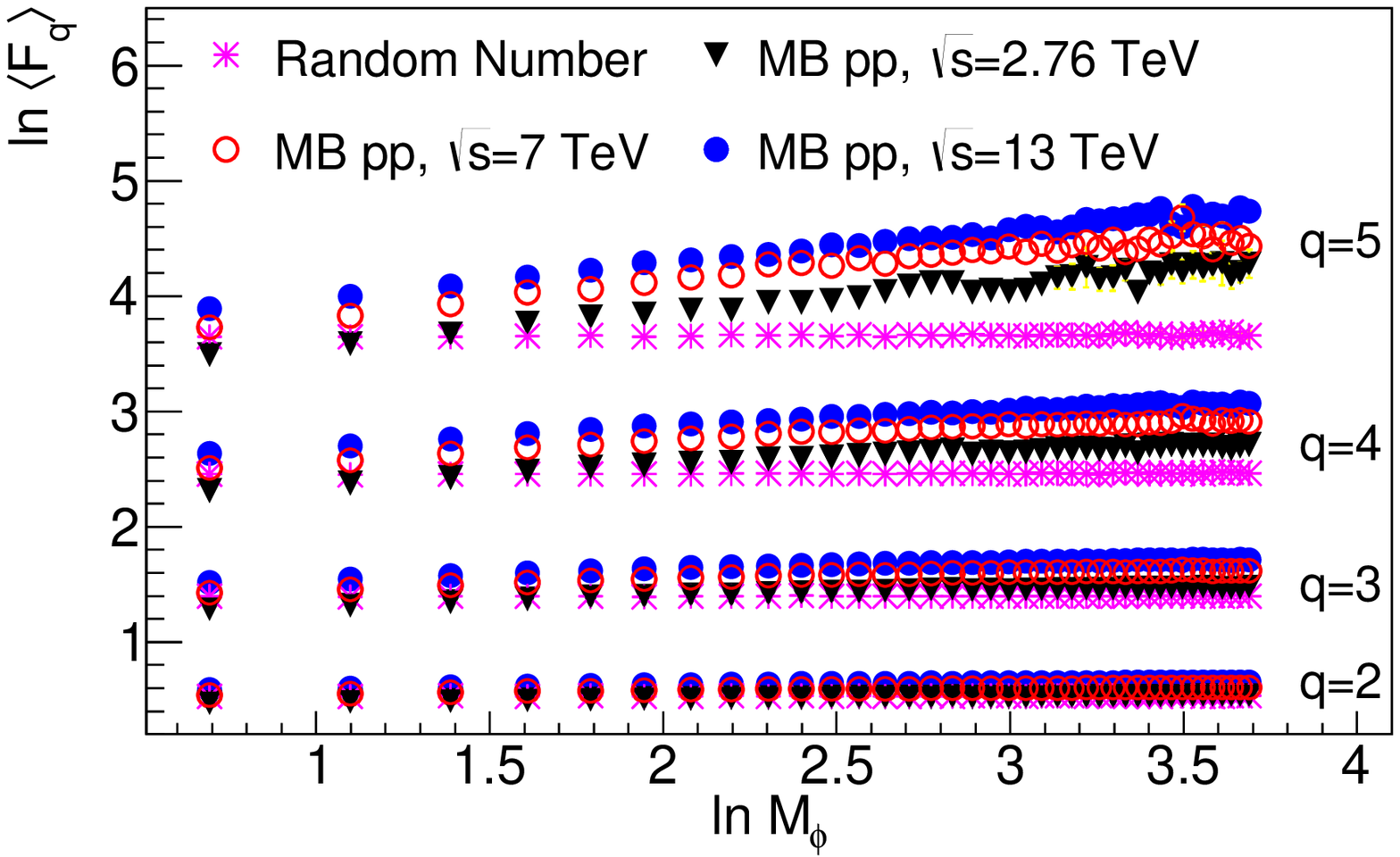}\label{1d_phi_sfm_mb190}}
\subfigure[]{\includegraphics[width=118mm, height=50mm]{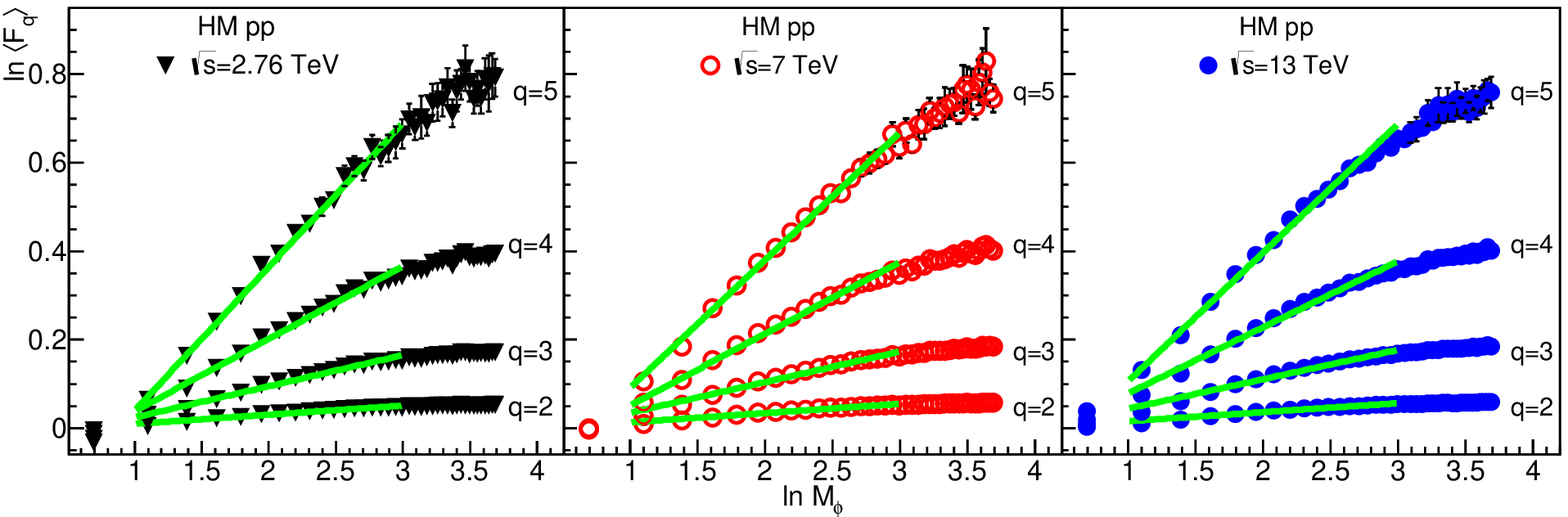}\label{1d_phi_sfm_hm215}}
\caption{(Color online) ln$\langle F_{q} \rangle$ vs. ln$M$ for moments $q=2-5$ for (a) RAN, MB $pp$ collisions and (b) HM $pp$ collisions at $\sqrt{s}=$ 2.76, 7, and 13 TeV in one-dimensional $\chi(\phi)$ space with PYTHIA Monash (default) generated data.}
\end{figure*}

Horizontally averaged scaled factorial moments $\langle F_{q} \rangle$ for different order $q=2-5$ for minimum-bias $pp$ collisions at $\sqrt{s}=$  2.76, 7, and 13 TeV have been estimated for $\chi(\eta)$ space using Eq. (\ref{HSFM_equation}) and plotted against the number of phase-space bins $M$ in log-log scale and is shown in Fig.~\ref{sfm_mb190}. An equal number of events are generated using a random number (RAN) generator with values lying between 0 and 1 and the ln$\langle F_{q} \rangle$ vs. ln$M$ is plotted in the same Fig.~\ref{sfm_mb190}. No significant rise in ln$\langle F_{q} \rangle$ against ln$M$ could be seen for both MB and the random number generated data giving no indication of the presence of any dynamical fluctuation in the emission spectra of primary particles of MB events in $\chi(\eta)$ space.

\begin{figure}[hb]
\includegraphics[width=65mm, height=54mm]{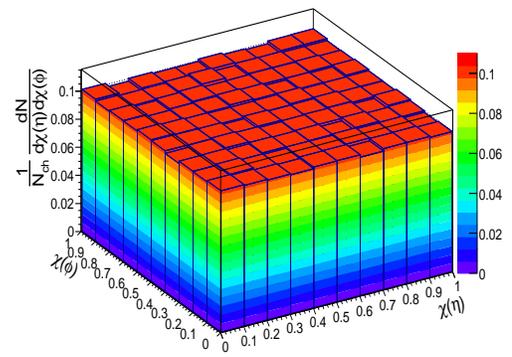}
\caption{(Color online) $\chi(\eta-\phi)$ distribution of the primary charged particles produced in HM $pp$ events at $\sqrt{s}$ = 13 TeV with PYTHIA Monash (default) generated data.}
\label{2dchi_dist}
\end{figure}

\begin{figure*}[htp]
\subfigure[]{\includegraphics[width=60mm, height=50mm]{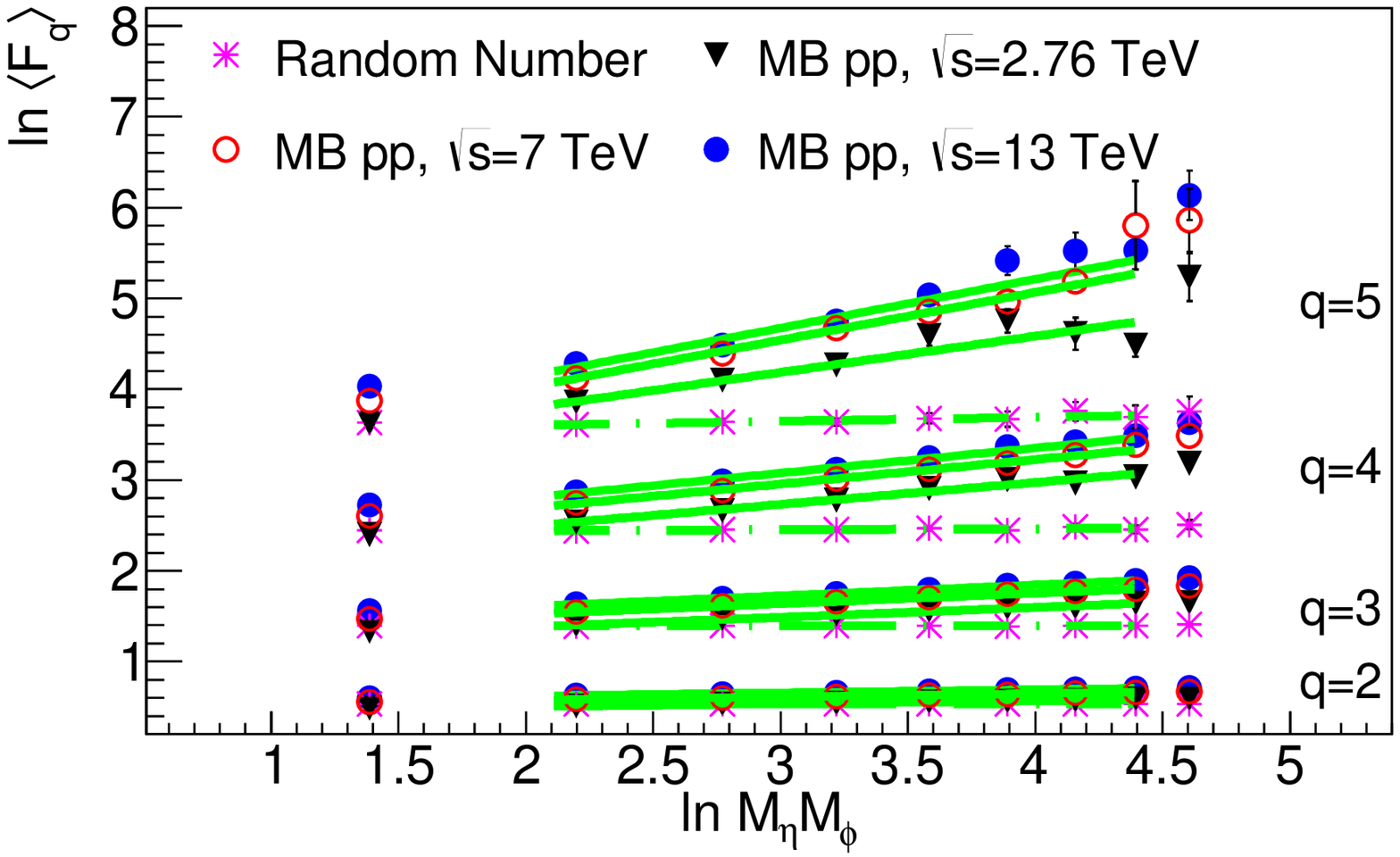}\label{2d_sfm_mb190}}
\subfigure[]{\includegraphics[width=118mm, height=50mm]{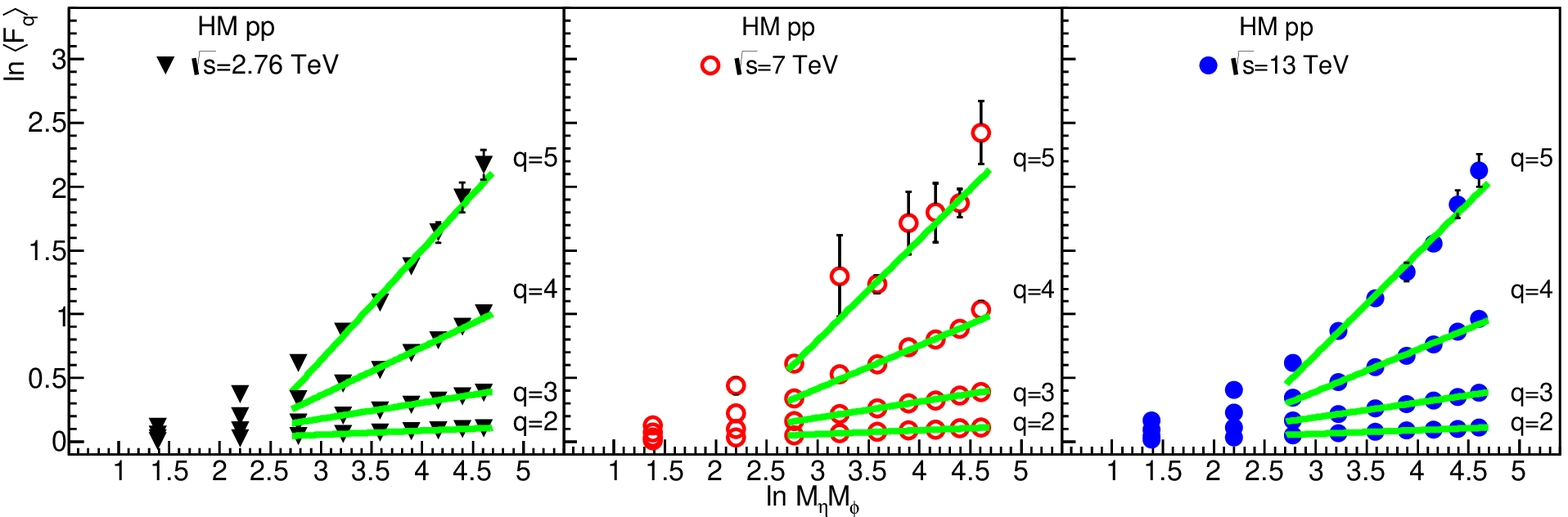}\label{2d_sfm_hm215}}
\caption{(Color online) ln$\langle F_{q} \rangle$ vs. ln$M$ for moments $q=2-5$ for (a) RAN, MB $pp$ collisions and (b) HM $pp$ collisions at $\sqrt{s}=$ 2.76, 7, and 13 TeV in two-dimensional $\chi(\eta-\phi)$ space with PYTHIA Monash (default) generated data.}
\end{figure*}

In Fig.~\ref{sfm_hm190}, the same plot is shown for HM $pp$ events at $\sqrt{s}=$ 2.76, 7, and 13 TeV. The green solid lines represent the straight line fit of the data points. The straight line fitting is done keeping the correlation coefficient value $R^2=$ 0.99. The errors shown in these plots are statistical errors only. It could be readily seen from the plot that ln$\langle F_{q} \rangle$ increases linearly with the increase of ln$M$ and the increase is more pronounced for the higher moments. This behavior is a clear indication of the presence of intermittency and multi-particle correlation in pseudorapidity space in the emission spectra of PYTHIA Monash (default) generated primary particles of high-multiplicity $pp$ events at $\sqrt{s}=$ 2.76, 7, and 13 TeV.

The intermittent pattern in the emission source of the particle in high-energy nuclear collisions may be different in different phase space, depending on the nature of the emission spectra. To realize  the (a)symmetric nature of the intermittent pattern in $pp$ collisions at the LHC energies, the same analysis has also been carried out in one-dimensional azimuthal ($\phi$) space.

The ln$\langle F_{q} \rangle$ against ln$M$ plots in one-dimensional $\chi(\phi)$ space for $q=2-5$ in minimum-bias and high-multiplicity $pp$ collisions at $\sqrt{s}=$ 2.76, 7, and 13 TeV are shown in Figs. \ref{1d_phi_sfm_mb190} and \ref{1d_phi_sfm_hm215} respectively. A small rise in ln$\langle F_{q} \rangle$ against ln$M$ is evident from Fig.~\ref{1d_phi_sfm_mb190} for moments $q=4$ and 5 in MB $pp$ collisions. On the other hand, from Fig.~\ref{1d_phi_sfm_hm215} for HM $pp$ events, a clear increase in the ln$\langle F_{q} \rangle$ against ln$M$ could be observed for $q=2-5$. The intermittency index $\alpha_{q}$, for the high-multiplicity $pp$ events of different energies, estimated from the straight line fit of the data points (ln$M$ = 1.0 to 3.0) in $\chi(\eta)$ and $\chi(\phi)$ spaces, are respectively found to be the same within the statistical error and therefore the values corresponding to $\sqrt{s}=$ 13 TeV are only listed in Table. \ref{phi_table}. From the obtained values of the $\alpha_q$, the emission of particles is found to be more intermittent in $\chi(\phi)$ space than in $\chi(\eta)$space.

It was pointed out by Ochs \cite{ochs2} that the intermittent behavior that occurs in higher dimensional space may disappear or saturate at small phase-space intervals in one-dimensional space. This may occur due to the reduction of fluctuation by the averaging process in the lower dimensional projection. Thus, the study of fluctuation in higher dimensional space is of significance, particularly for MB $pp$ collisions.

To study the fluctuation in two-dimensional pseudorapidity-azimuthal ($\eta-\phi)$ space, the one-dimensional $\chi(\eta)$ and $\chi(\phi)$ distributions have been mapped into a two-dimensional $\chi(\eta-\phi)$ distribution as shown in Fig. \ref{2dchi_dist}. The two-dimensional $\chi(\eta-\phi)$ space is now successively divided into $M_i \times M_i$ bins of equal width $\delta\chi_{\eta}\times\delta\chi_{\phi}$ where, $i=$ 1 to 10. The numbers of particles in each square bin are counted and its corresponding scaled factorial moments (SFMs) are estimated. Finally, the obtained SFMs are averaged over all bins and all events to obtain $\langle F_q \rangle$ for $M=M_i \times M_i$ = 2$\times$2 = 4, for example \cite{adamovich, mali}.

\begin{table}[h]
\caption{Values of the intermittency index $\alpha_{q}$ in high-multiplicity $pp$ events at $\sqrt{s}=$ 13 TeV for different moments $q$ in $\chi(\eta)$ and $\chi(\phi)$ spaces.}
\label{phi_table}
\begin{ruledtabular}
\newcolumntype{C}[1]{>{\centering\arraybackslash}p{#1}}
\renewcommand\familydefault{\sfdefault}
{\def\arraystretch{1.2}
\begin{tabular}{c c c c c c c}
\multirow{2}{*}{$q$} &\multicolumn{2}{c}{Intermittency index ($\alpha_{q}$) for $\sqrt{s}=$ 13 TeV} & \multirow{2}{*} {$R^{2}$} {}\\
  &  $\chi(\eta)$ space &   $\chi(\phi)$ space   & \\
  \hline
 2   &   0.0022 $\pm$ 0.0004		&  0.0202 $\pm$ 0.0005   &   	0.99		\\
 3   &   0.0070 $\pm$ 0.0008 	&   0.066 $\pm$ 0.0010   &    	0.99		\\
 4   &   0.016 $\pm$ 0.0014  	&	0.150 $\pm$ 0.0019   &      0.99		\\
 5   &   0.031 $\pm$ 0.0029   	&   0.291 $\pm$ 0.0038	 &    	0.99 	\\
\end{tabular}}
\end{ruledtabular}
\end{table}

\begin{table}[h]
\caption{Values of the intermittency index $\alpha_{q}$ in minimum-bias and high-multiplicity $pp$ collisions at $\sqrt{s}=$ 13 TeV for different moments $q$ in $\chi(\eta-\phi)$ space.}
\label{phi_table1}
\begin{ruledtabular}
\centering
\newcolumntype{C}[1]{>{\centering\arraybackslash}p{#1}}
\renewcommand\familydefault{\sfdefault}
{\def\arraystretch{1.2}
\begin{tabular}{c c c c c}
\multirow{2}{*}{$q$} &\multicolumn{2}{c}{Intermittency index ($\alpha_{q}$) for $\sqrt{s}=$ 13 TeV} & \multirow{2}{*} {$R^{2}$} {}\\
&     \multicolumn{2}{c}{$\chi(\eta-\phi)$ space}	& \\
&  Minimum bias  & High multiplicity \\
  \hline
 2   &   0.034 $\pm$ 0.002	& 0.029$\pm$0.001	 	& 	0.99		\\
 3   &   0.119 $\pm$ 0.004 	& 0.116$\pm$0.002		& 	0.99		\\
 4   &   0.276 $\pm$ 0.012  	& 0.330$\pm$0.008		&   0.99		\\
 5   &   0.538 $\pm$ 0.034  & 0.794$\pm$0.027 		& 	0.99 	\\
\end{tabular}}
\end{ruledtabular}
\end{table}

The ln$\langle F_{q} \rangle$ against ln$M$ plots in two-dimensional $\chi(\eta-\phi)$ space for $q=2-5$ for minimum-bias and high-multiplicity $pp$ collisions at $\sqrt{s}=$ 2.76, 7, and 13 TeV are shown in Figs.~\ref{2d_sfm_mb190} and \ref{2d_sfm_hm215} respectively. From the two-dimensional analysis, a weak intermittent type of emission could now be seen for MB $pp$ events for all the studied energies as shown in Fig.~\ref{2d_sfm_mb190}. For all three energies, the intermittency index $\alpha_{q}$, within the statistical error, are respectively found to be the same for MB and HM $pp$ events and therefore, the values corresponding to $\sqrt{s}=$ 13 TeV are only listed in Table. \ref{phi_table1}.

\begin{figure}[htp]
\includegraphics[width=75mm,height=57mm]{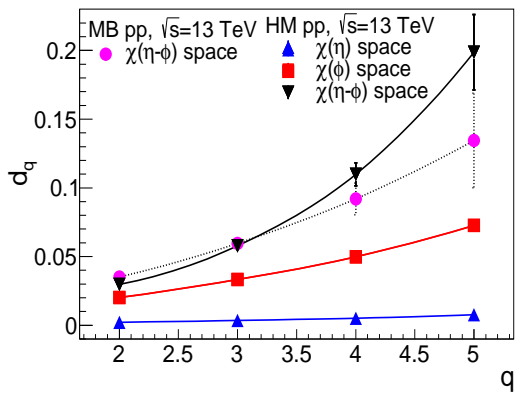}
\caption{(Color online) Variation of  $d_{q}$ against $q$ for MB [$\chi(\eta-\phi)$ space] and for HM [$\chi(\eta)$, $\chi(\phi)$, and $\chi(\eta-\phi)$ spaces] $pp$ event at $\sqrt{s}=$ 13 TeV with PYTHIA Monash (default) generated data.}
\label{dq_all}
\end{figure}

The anomalous dimension $d_{q}$ has been estimated using the Eq. (\ref{dq_equation}) for MB $pp$ events in $\chi(\eta-\phi)$ space and for high-multiplicity $pp$ events in $\chi(\eta)$, $\chi(\phi)$, and $\chi(\eta-\phi)$ spaces at $\sqrt{s}=$ 13 TeV. The variations of $d_{q}$ with the order of the moments $q$ are shown in Fig.~\ref{dq_all}. The dotted (MB) and solid (HM) lines in the figure are drawn to guide the eyes only. From the figure, an increase in $d_{q}$ with the order of the moment $q$ could be observed for MB $pp$ events in the two-dimensional $\chi(\eta-\phi)$ space and for high-multiplicity $pp$ events in all the $\chi(\eta)$, $\chi(\phi)$, and $\chi(\eta-\phi)$  spaces. The increase is found to be more pronounced in $\chi(\eta-\phi)$ space than that of $\chi(\eta)$, and $\chi(\phi)$ spaces. Such an increase of $d_{q}$ with $q$ indicates the presence of multifractal behavior in the emission spectra of the particles of MB and HM $pp$ events, which could be attributed to a cascading mechanism of particles production in such $pp$ collisions \cite{bialas2, ochs1, bialas1}.

Bialas and Zalewski \cite{bialas4} reported that the intermittent behavior in the final-state particles in the ultra-relativistic collisions may also be a projection of non-thermal phase transition that occurs during the evolution of the collisions which in turn would be responsible for the occurrence of the  anomalous events. The presence of a non-thermal phase transition is expected to have a minimum value of coefficient $\lambda_{q}$ at some value of $q=q_{c}$, where $\lambda_{q}$ is related to $\alpha_{q}$ through the relation

\begin{equation}
\lambda_{q}=\frac{\alpha_{q}+1}{q}.
\label{lambda_q}
\end{equation}

The value of $q_{c}$  need not  necessarily be an integer and the region satisfying the condition $q<q_{c}$ may be dominated by many small fluctuations; whereas the region $q>q_{c}$ contains rarely occurring large fluctuations.

\begin{figure}[h]
\includegraphics[width=75mm,height=57mm]{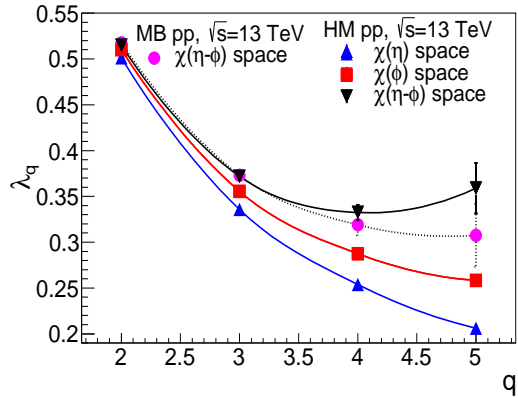}
\caption{(Color online) Variation of $\lambda_{q}$ against $q$ for MB [$\chi(\eta-\phi)$ space] and for HM [$\chi(\eta)$, $\chi(\phi)$, and $\chi(\eta-\phi)$ spaces] $pp$ event at $\sqrt{s}=$ 13 TeV with PYTHIA Monash (default) generated data.}
\label{lambda_all}
\end{figure}

The variations of $\lambda_{q}$ against $q$ in different spaces for MB [$\chi(\eta-\phi)$] and HM [$\chi(\eta)$, $\chi(\phi)$, and $\chi(\eta-\phi)$] $pp$ events at $\sqrt{s}=$ 13 TeV are shown in Fig.~\ref{lambda_all}. $\lambda_{q}$ is found to decrease monotonically with $q$ for MB $pp$ events, thereby ruling out any possibility of occurrence of nonthermal phase-transition-like behavior in such collisions. However, it is interesting to note for HM $pp$ events that, though the $\lambda_{q}$ value for one-dimensional $\chi(\eta)$ and $\chi(\phi)$ spaces decreases monotonically with $q$, a clear minimum in $\lambda_{q}$, estimated from two-dimensional analysis, could be observed at $q=q_c=4$. The observation of a clear minimum in $\lambda_{q}$ at $q=q_c=4$ is indicative of a nonthermal phase-transition-like behavior in high-multiplicity $pp$ events.

\begin{figure*}[ht]
\subfigure[$\chi(\eta)$ space]{\includegraphics[width=59mm,height=50mm]{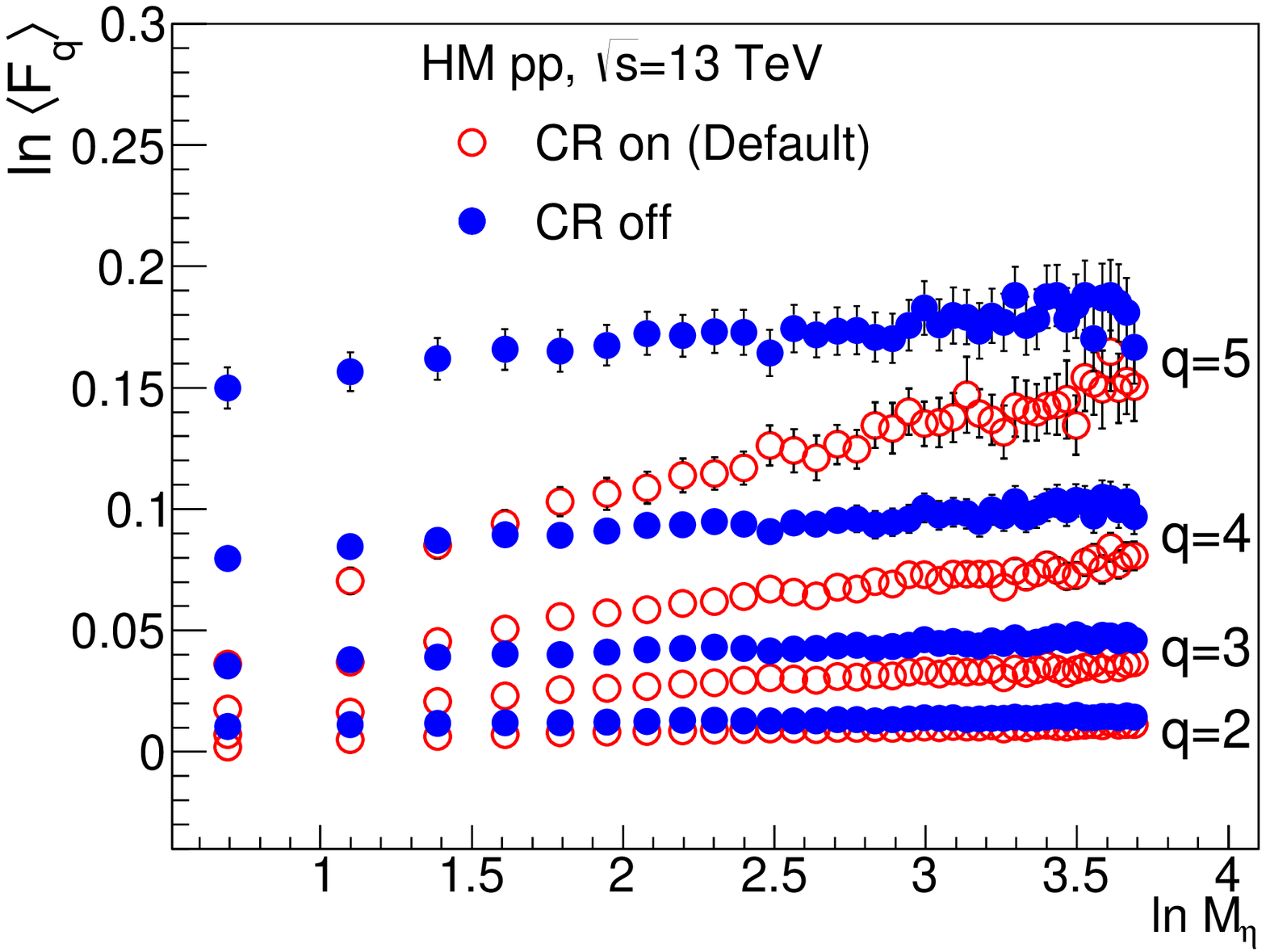}\label{cr_off_eta}}
\subfigure[$\chi(\phi)$ space]{\includegraphics[width=59mm,height=50mm]{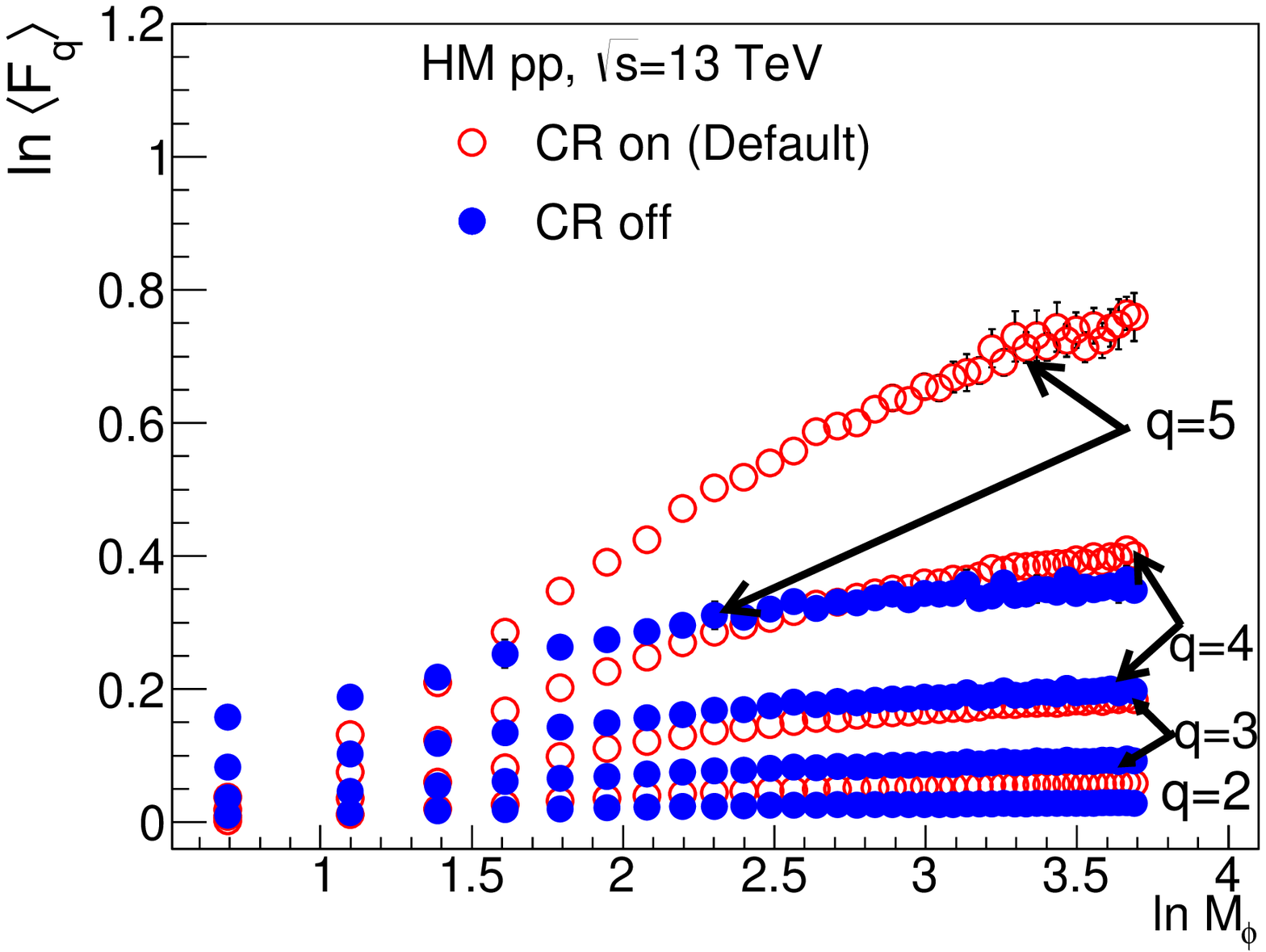}\label{cr_off_phi}}
\subfigure[$\chi(\eta-\phi)$ space]{\includegraphics[width=59mm,height=50mm]{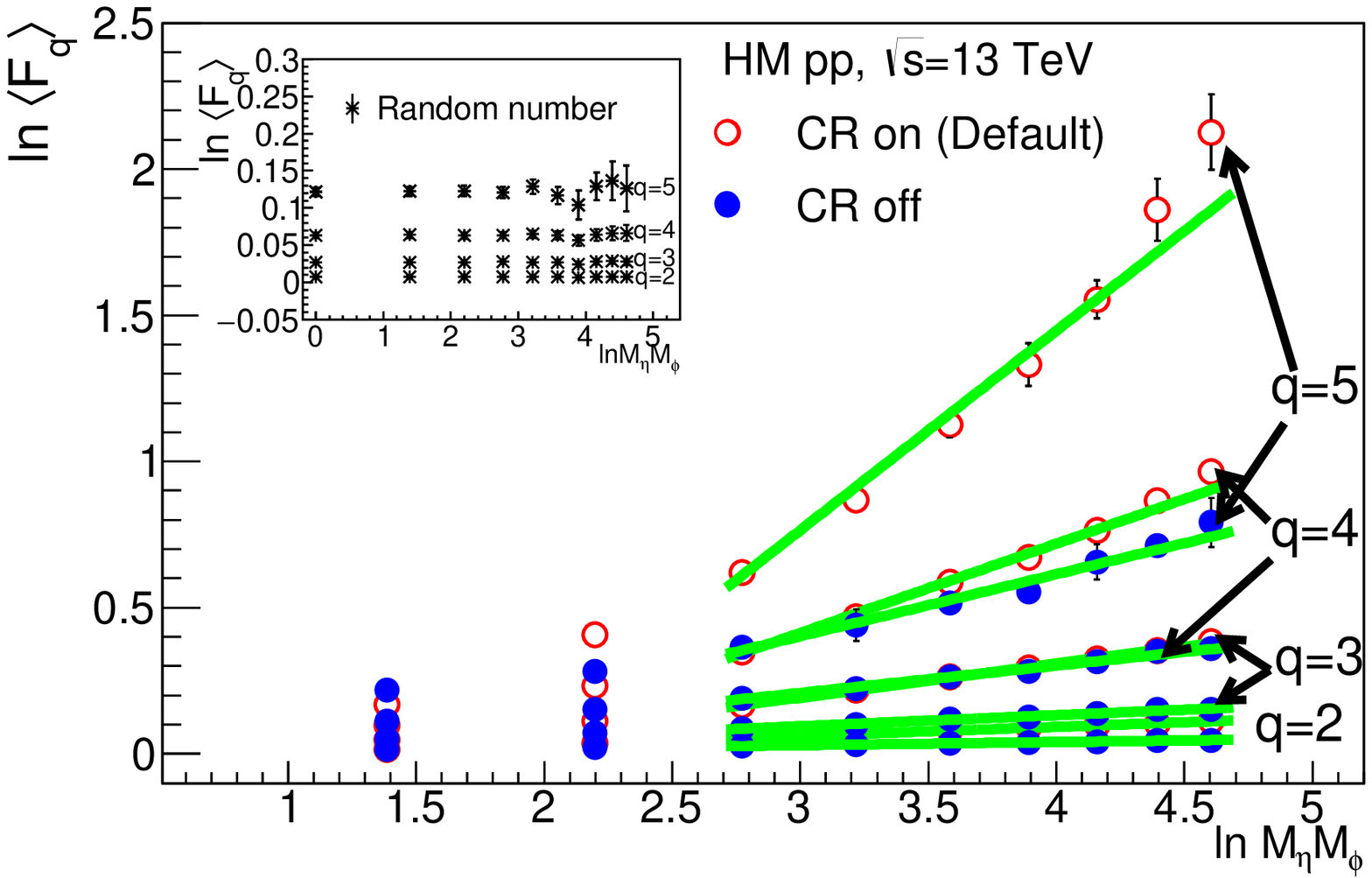}\label{cr_off_etaphi}}
\caption{(Color online) ln$\langle F_{q} \rangle$ vs. ln$M$ for moments $q=2-5$ for HM $pp$ events at $\sqrt{s}=$ 13 TeV in (a) $\chi(\eta)$, (b) $\chi(\phi)$, and (c) $\chi(\eta-\phi)$ spaces for PYTHIA Monash generated data with CR on (RR = 1.8) and CR off (RR = 0.0). Inset plot in (c) shows the ln$\langle F_{q} \rangle$ vs. ln$M$ for random number.}
\end{figure*}

\begin{figure*}[htp]
\subfigure[$\chi(\eta)$ space]{\includegraphics[width=59mm,height=50mm]{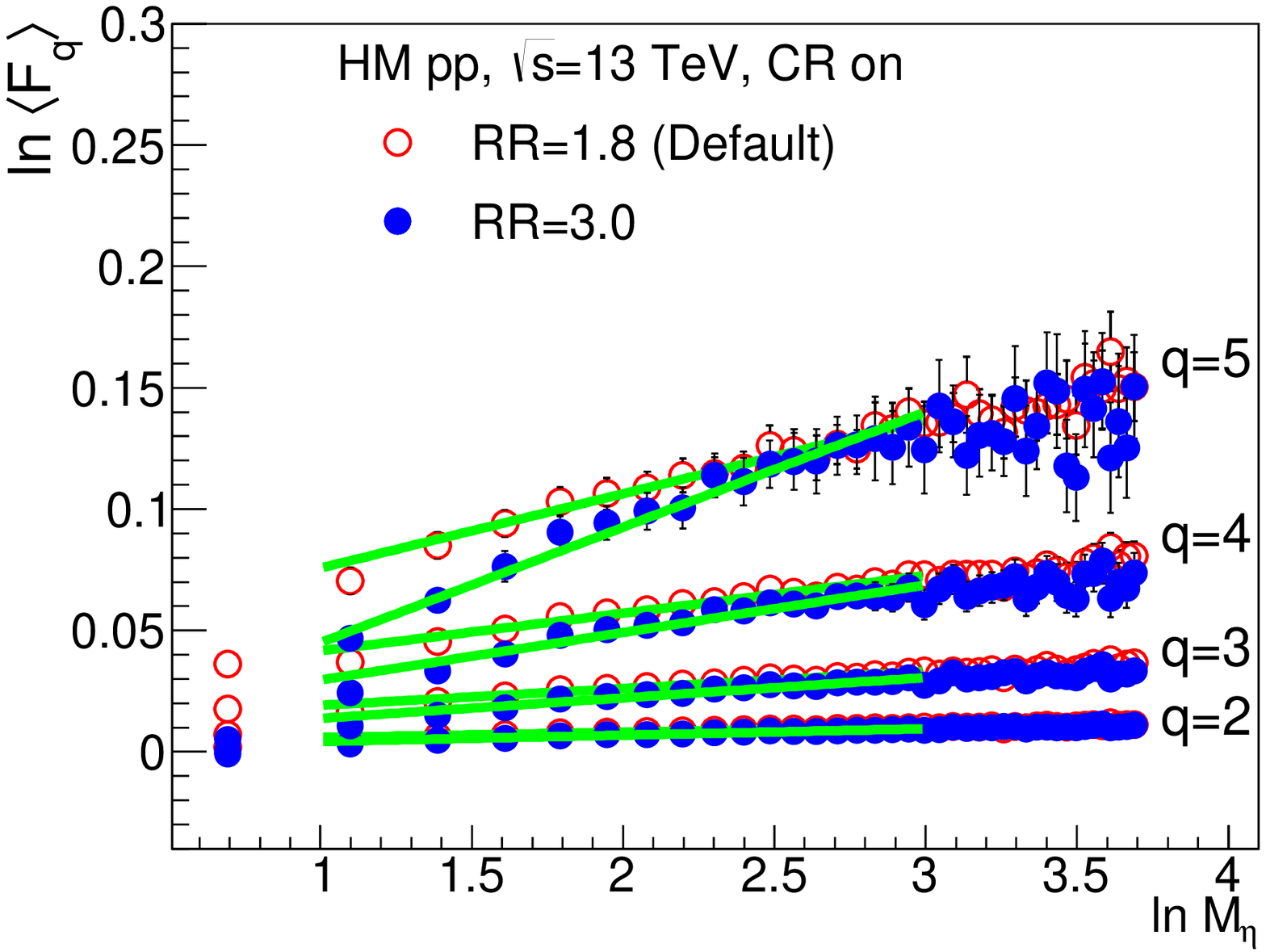}\label{cr_3_eta}}
\subfigure[$\chi(\phi)$ space]{\includegraphics[width=59mm,height=50mm]{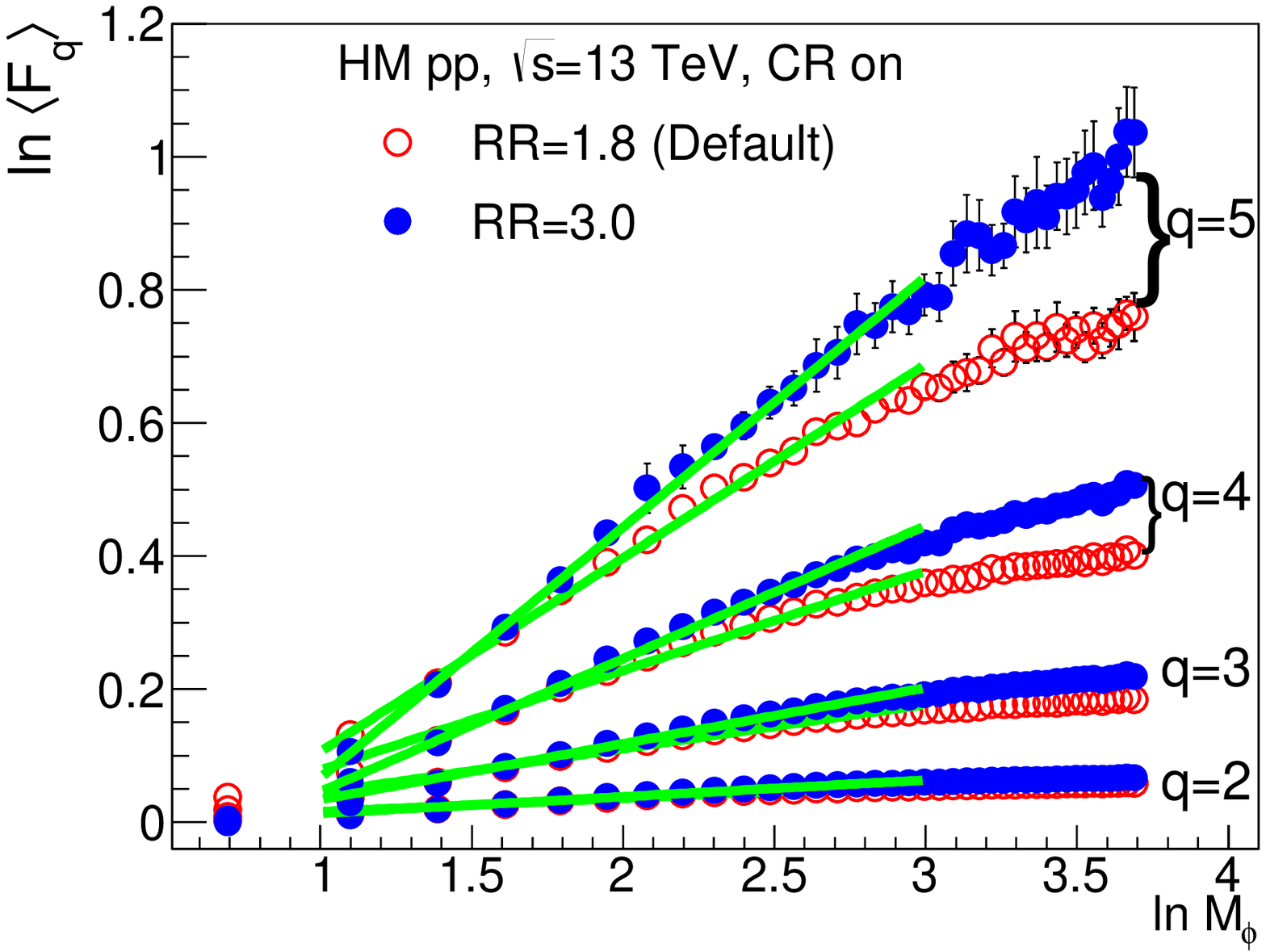}\label{cr_3_phi}}
\subfigure[$\chi(\eta-\phi)$ space]{\includegraphics[width=59mm,height=50mm]{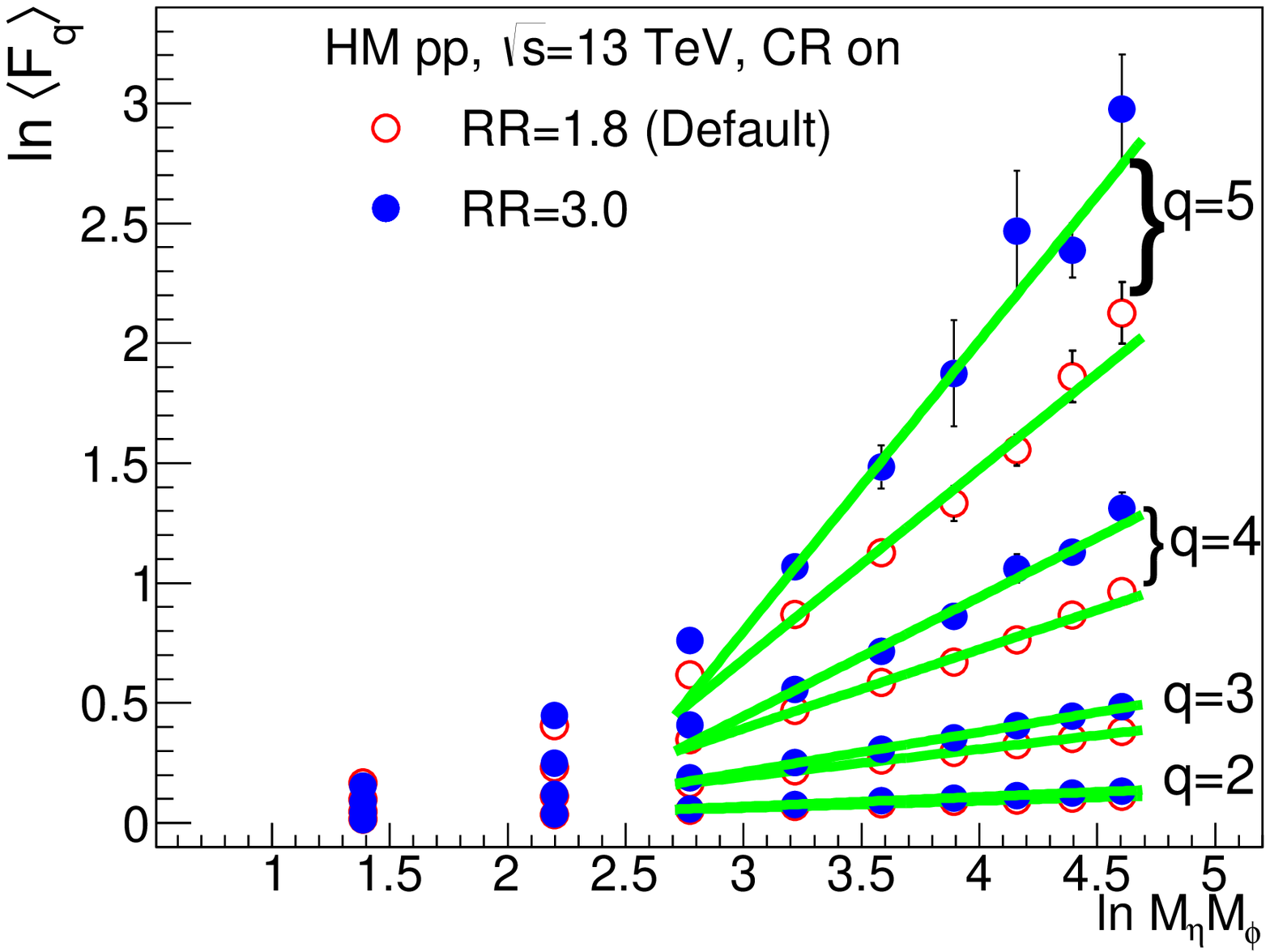}\label{cr_3_etaphi}}
\caption{(Color online) ln$\langle F_{q} \rangle$ vs. ln$M$ for moments $q=2-5$ for HM $pp$ events at $\sqrt{s}=$ 13 TeV in (a) $\chi(\eta)$, (b) $\chi(\phi)$, and (c) $\chi(\eta-\phi)$ spaces for PYTHIA Monash generated data with RR = 1.8 and 3.0.}
\end{figure*}

\begin{figure*}[ht]
\subfigure[$d_{q}$ vs. q]{\includegraphics[width=75mm,height=58mm]{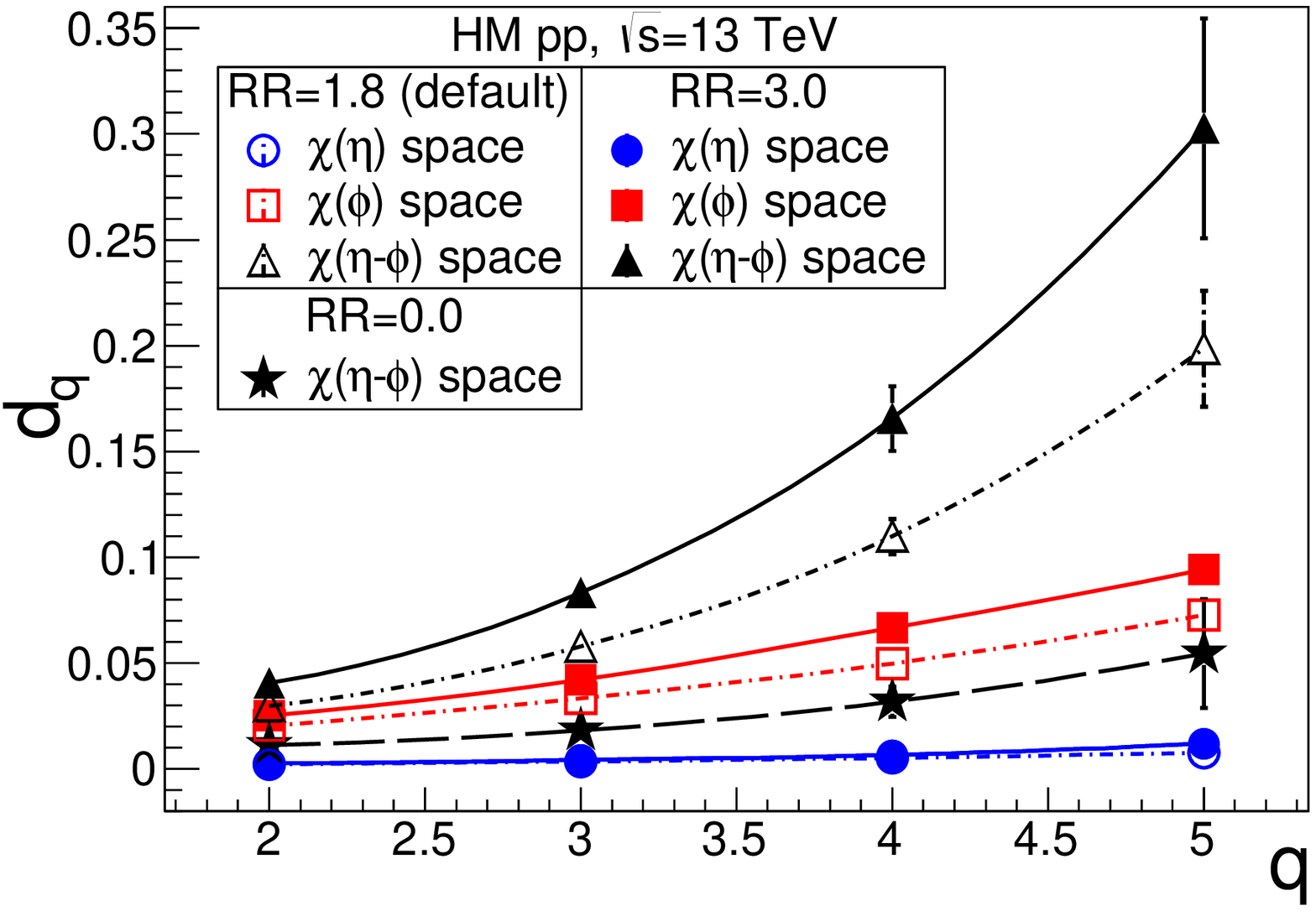}\label{dq_rr3}}
\subfigure[$\lambda_{q}$ vs. q]{\includegraphics[width=75mm,height=58mm]{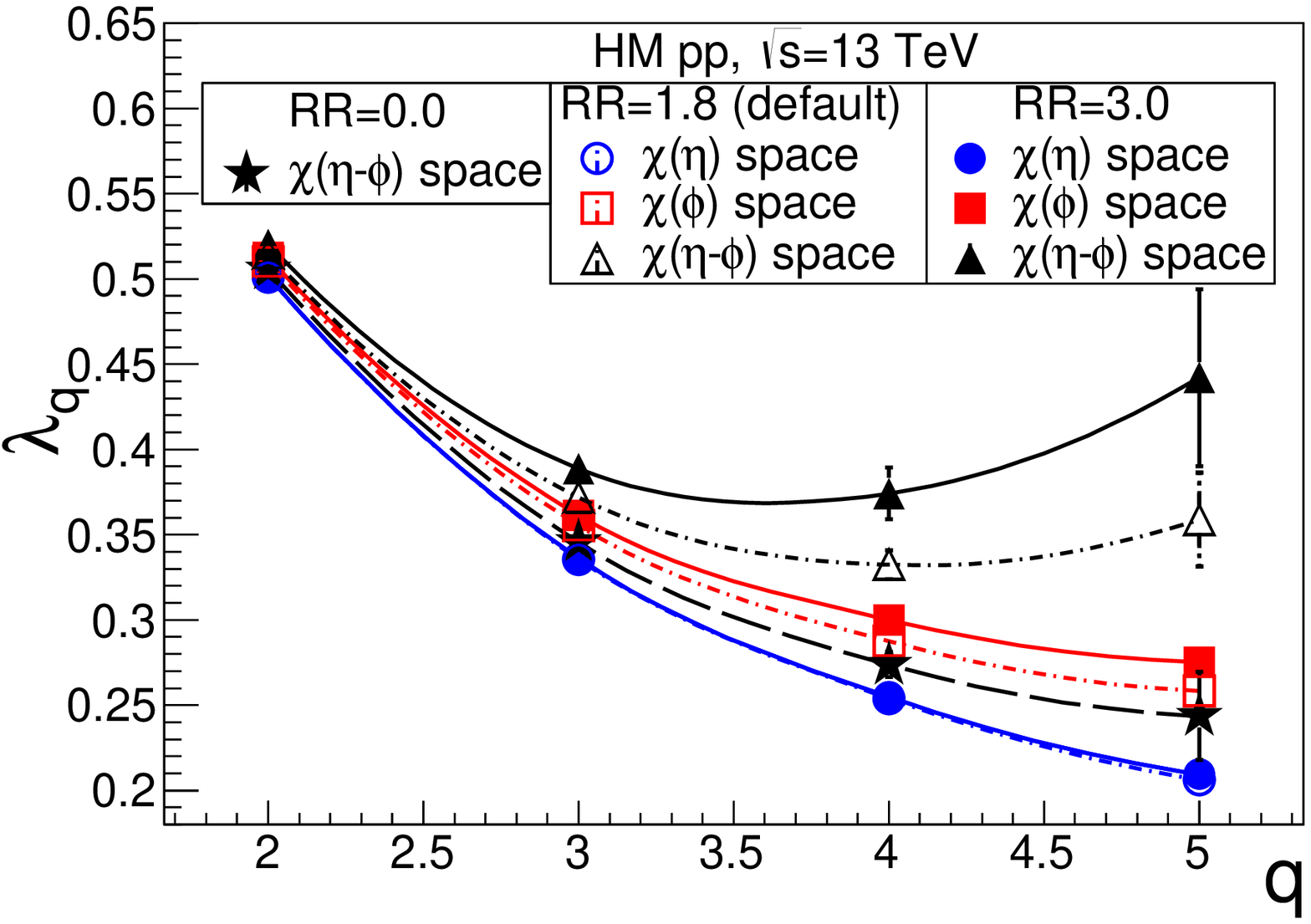}\label{lambda_rr3}}
\caption{(Color online) Variation of (a) $d_{q}$ against $q$ and (b) $\lambda_{q}$ against $q$ for HM $pp$ event at $\sqrt{s}=$ 13 TeV for PYTHIA Monash generated data with RR = 0.0 in $\chi(\eta-\phi)$ space and with RR = 1.8 and 3.0 in $\chi(\eta)$, $\chi(\phi)$, and $\chi(\eta-\phi)$ spaces.}
\end{figure*}

\begin{figure*}[ht]
\subfigure[$\chi(\eta)$ space]{\includegraphics[width=59mm,height=50mm]{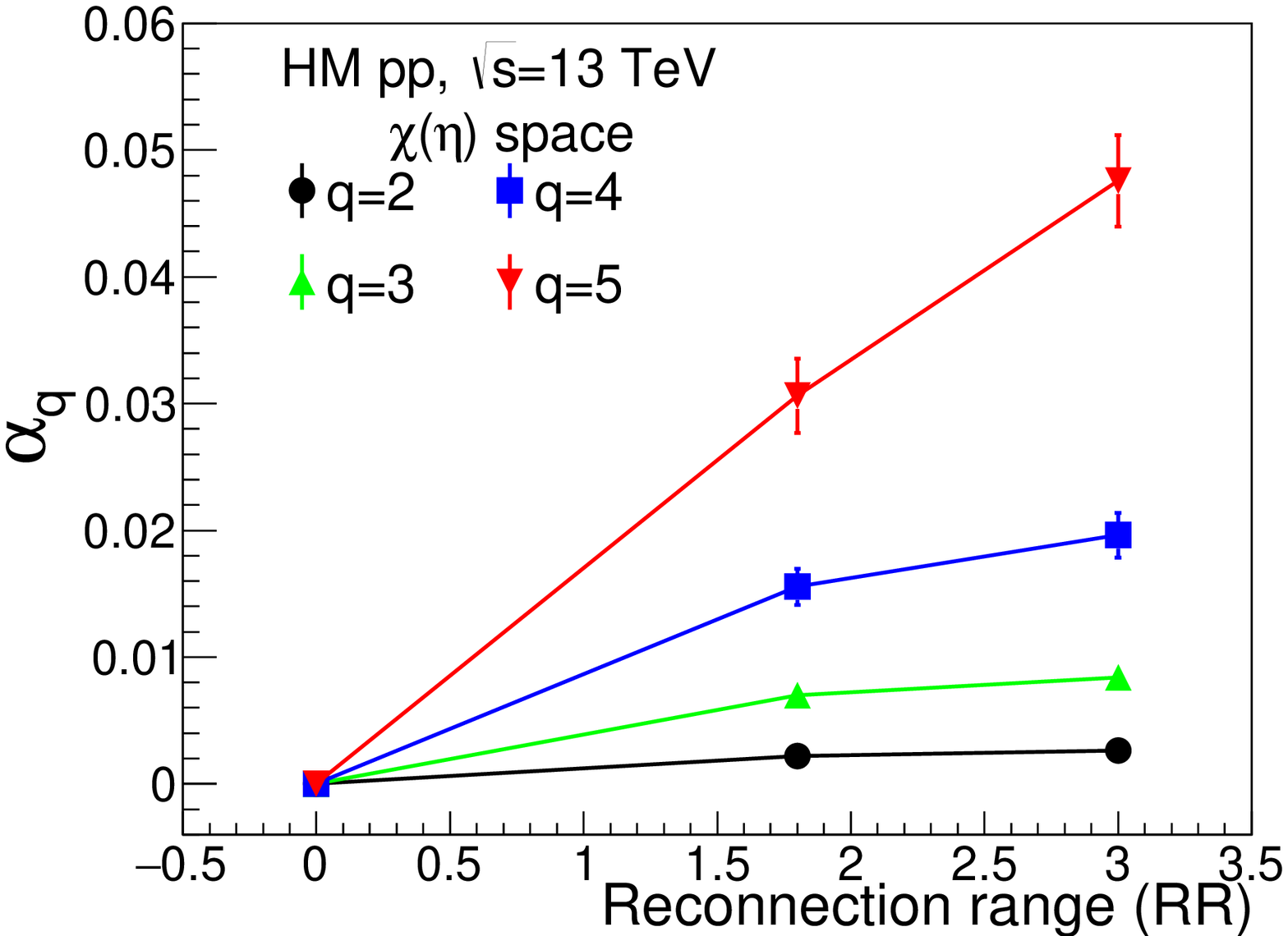}\label{rr_comp_eta}}
\subfigure[$\chi(\phi)$ space]{\includegraphics[width=59mm,height=50mm]{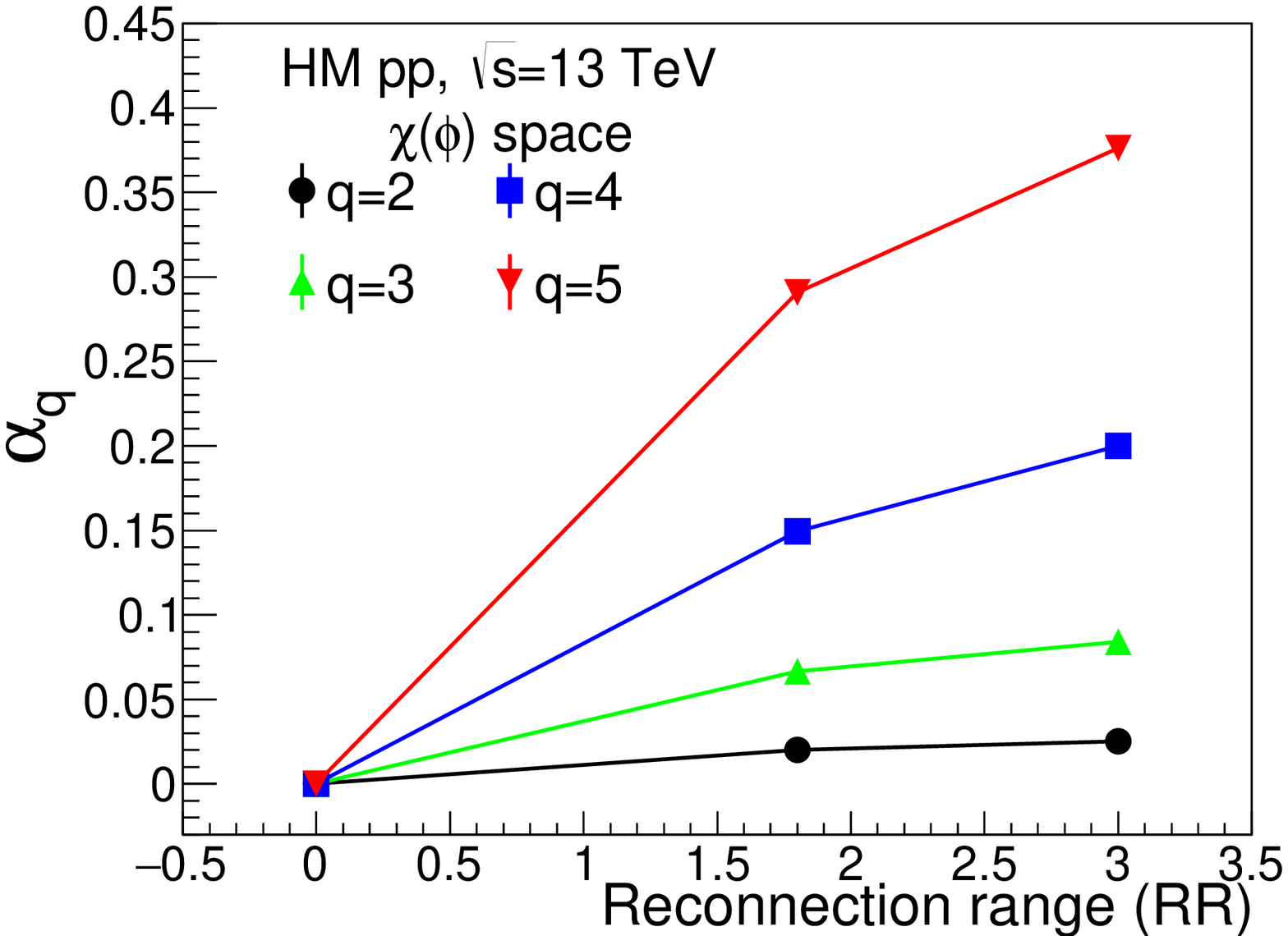}\label{rr_comp_phi}}
\subfigure[$\chi(\eta-\phi)$ space]{\includegraphics[width=59mm,height=50mm]{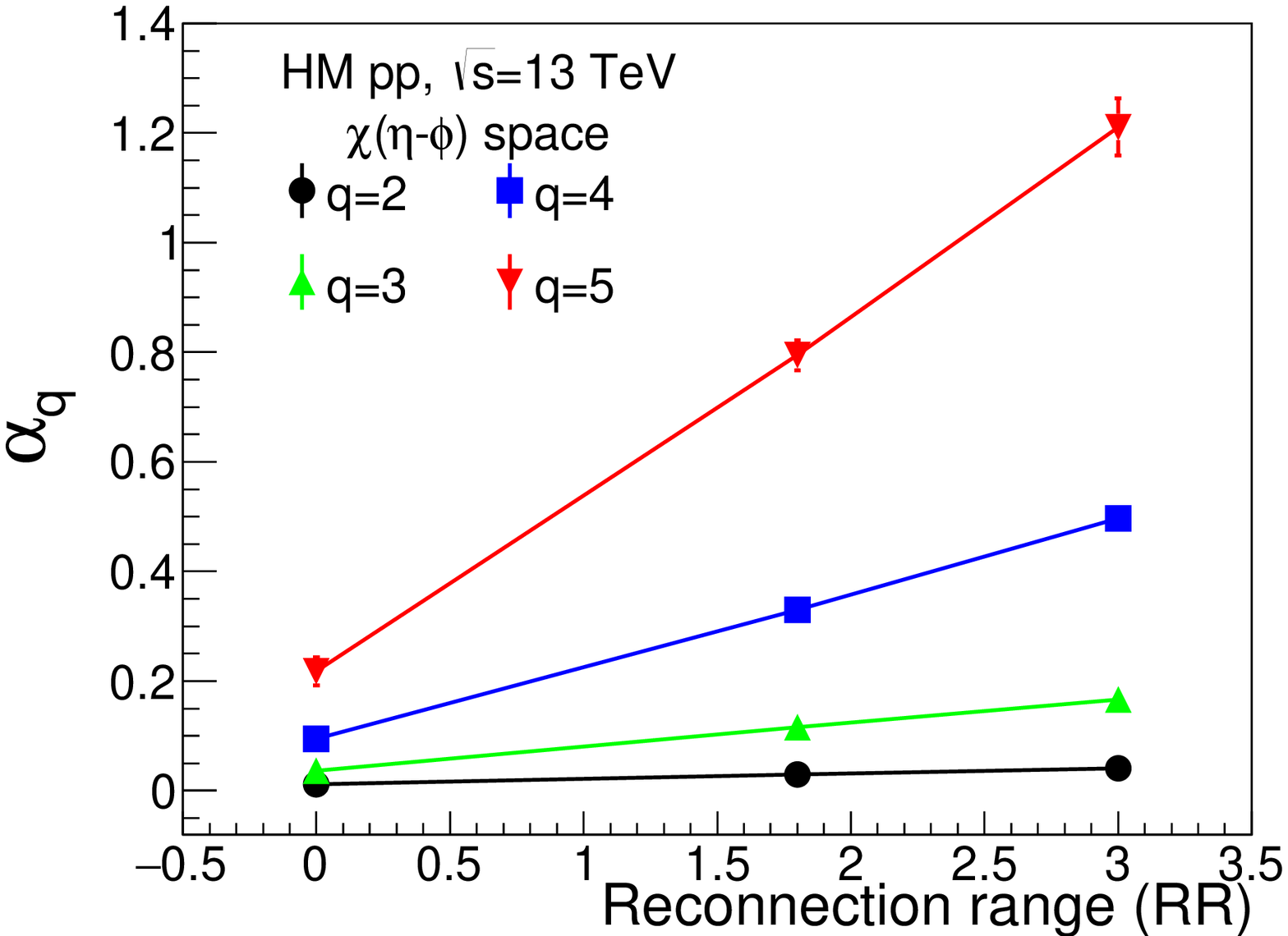}\label{rr_comp_etaphi}}
\caption{(Color online) Variation of intermittency index $\phi_{q}$ against reconnection range (RR) for $q=2-5$ in (a) $\chi(\eta)$, (b) $\chi(\phi)$, and (c) $\chi(\eta-\phi)$ spaces in HM PYTHIA Monash generated $pp$ events at $\sqrt{s}=$ 13 TeV.}
\end{figure*}

In PYTHIA, color reconnection (CR) is a string fragmentation model where final partons are considered to be color connected in such a way that the total string length becomes as short as possible~\citep{gustafson}. The fragmentation of two independent interactions selects a preferred pseudorapidity minimizing $\Delta\eta$. Such an effect might give rise to large fluctuation in narrow pseudorapidity space.

Color reconnection in PYTHIA is introduced through a parameter called reconnection range (RR). In the default PYTHIA Monash model RR is taken to be equal to 1.8. To investigate the effect of color reconnection in our scaled factorial moments estimation in $pp$ data, a new set of (16.6 $\times$ $10^6$) $pp$ events has been generated by switching off (RR = 0.0) the color reconnection mechanism at $\sqrt{s}=$ 2.76, 7, and 13 TeV and the same analyses have been carried out for the high-multiplicity $pp$ events. For such events, as the values of the observed  intermittency index and other derived quantities, within statistical error, are found to be the same in respective spaces for $\sqrt{s}=$ 2.76, 7, and 13 TeV, the results of the highest studied energy will only be discussed hereafter. The obtained results for ln$\langle F_{q} \rangle$ against ln$M$ in $\chi(\eta)$, $\chi(\phi)$ and $\chi(\eta-\phi)$ spaces in high-multiplicity $pp$ events at $\sqrt{s}=$ 13 TeV are shown Figs.~\ref{cr_off_eta}, \ref{cr_off_phi}, and \ref{cr_off_etaphi} respectively and compared with the results of our default (RR = 1.8) PYTHIA data. A significant decrease in the strength of the intermittency could readily be observed from these figures for moments $q=2-5$ for CR off data set.

From the above observation, it is evident that color reconnection plays a significant role in the observed intermittency in the PYTHIA Monash (default) generated data set at $\sqrt{s}=$ 2.76, 7, and 13 TeV. For further confirmation of the effect of CR on the observed intermittent behavior, the RR parameter has been changed to 3.0 and a new set of 43.5 $\times$ $10^6$ data has been generated again for $pp$ collisions at $\sqrt{s}=$ 13 TeV.

The ln$\langle F_{q} \rangle$ against ln$M$ plots for different moments for high-multiplicity $pp$ collisions at $\sqrt{s}=$ 13 TeV with RR = 3.0 are shown in Figs.~\ref{cr_3_eta}, \ref{cr_3_phi}, and \ref{cr_3_etaphi} for $\chi(\eta)$, $\chi(\phi)$, and $\chi(\eta-\phi)$ spaces, respectively, and compared with the results of default (RR = 1.8) PYTHIA Monash generated data set. Here also, the green solid lines represent the straight line fit of the data points and the fitting is done keeping $R^2=$ 0.99. From these figures, a sharp increase in the values of ln$\langle F_{q} \rangle$ against ln$M$ is clearly evident for RR = 3.0 than that of the default one. This behavior confirms that the color reconnection mechanism plays a significant role in the observed intermittent type of emission of primary charged particles of our PYTHIA Monash generated sets of data of $pp$ collisions.

The anomalous dimension $d_{q}$ and the coefficient $\lambda_{q}$, estimated using Eqs. (\ref{dq_equation}) and (\ref{lambda_q}) respectively for RR = 3.0 for $\chi(\eta)$, $\chi(\phi)$, and $\chi(\eta-\phi)$ spaces are plotted against $q$ in Figs.~\ref{dq_rr3},~\ref{lambda_rr3}, respectively, and compared with the estimated values with RR = 0.0 and 1.8 (default) PYTHIA Monash generated data. Though, a significant change in the values of $d_{q}$ could be observed with the order of the moments $q$ (Fig.~\ref{dq_rr3}) for RR = 1.8 and 3.0, not much change in $d_q$ with $q$ could be observed for RR = 0.0. Further, it is interesting to note from Fig.~\ref{lambda_rr3} that, whereas for RR = 0.0 data, no minimum could be seen in the $\lambda_q$ vs. $q$ plot, the minimum in $\lambda_q$ is found to be shifted towards lower values of $q=q_c=3.65$ for RR = 3.0. Such behavior confirms that for data set RR = 1.8 (default) and RR = 3.0 a nonthermal phase-transition-like behavior is conspicuous in the studied $pp$ events.

\section{Summary}

The intermittent pattern in the emission spectra of the primary charged particles produced in high-multiplicity $pp$ events at the LHC energies $\sqrt{s}=$ 2.76, 7, and 13 TeV could be seen in one-dimensional pseudorapidity ($\eta$), azimuthal ($\phi$), and two-dimensional pseudorapidity-azimuthal ($\eta-\phi$) spaces with the PYTHIA Monash (default) generated MC data. On the other hand, little ($\phi$, $\eta-\phi$ spaces) or no ($\eta$ space) such signature of intermittency could be seen with the same sets of data for minimum-bias $pp$ collisions. In the high-multiplicity $pp$ events, the intermittency index $\alpha_{q}$ increases with the increase of the order of the moments $q$ in all the three $\chi(\eta)$, $\chi(\phi)$, and $\chi(\eta-\phi)$ spaces. Further, the values of $\alpha_{q}$ for the various order of the moment $q$ is found to be most in $\chi(\eta-\phi)$ space and least in $\chi(\eta)$ space. No center of mass energy dependence in the strength of intermittency index could be seen for the studied systems ($pp$ at $\sqrt{s}=$ 2.76, 7, and 13 TeV). Estimation of anomalous dimension $d_{q}$ and its variation with the order of the moment $q$ suggests a multifractal nature of emission spectra of high-multiplicity $pp$ events and is attributed to the particle production through cascading mechanism. The coefficient $\lambda_q$ decreases monotonically with the increase of $q$ and no minimum value of $\lambda_{q}$ is evident in the $\lambda_{q}$ vs. $q$ plot in $\chi(\eta)$ and $\chi(\phi)$ spaces. On the other hand, a clear minimum value of $\lambda_{q}$ at $q=q_c=4$ is evident in $\chi(\eta-\phi)$ space and is indicative of the occurrence of nonthermal phase-transition-like behavior in the studied high-multiplicity $pp$ events. With color reconnection off (RR = 0.0) in PYTHIA, a small rise in ln$\langle F_{q} \rangle$ vs. ln$M$ for higher values of $q$ could be observed in high-multiplicity $pp$ events in $\chi(\eta-\phi)$ space. With the increase of reconnection range (RR), the controlling parameter of CR, a significant increase in the intermittent behavior could be observed in comparison to default (RR = 1.8) and RR = 0.0 PYTHIA data [Figs.~\ref{rr_comp_eta}, ~\ref{rr_comp_phi}, and ~\ref{rr_comp_etaphi}]. Moreover, the position of the $\lambda_{q}$ minimum is found to decrease ($q=q_c=3.65$) with the increase of color reconnection parameter (RR = 3.0). No such minimum in $\lambda_q$ could be seen with CR off (RR = 0.0). Thus, from this study, it is evident that the color reconnection mechanism in PYTHIA has a significant effect on the observed intermittency and hence on the nonthermal phase-transition-like behavior in the studied high-multiplicity events of $pp$ collisions.

\begin{acknowledgments}

The authors thank Prof. Rudolph C. Hwa of the University of Oregon, USA for his valuable comments and careful reading of the manuscript. P.S. thanks his colleague Nur Hussain for providing help in generating PYTHIA events. The authors thankfully acknowledge the Department of Science and Technology (DST), Government of India, for providing funds via Project No. SR/MF/PS-01/2014-GU(C) to develop a high-performance computing cluster (HPCC) facility for the generation of the Monte Carlo events for this work. 
\end{acknowledgments}

%

\end{document}